\documentclass[aps,preprint,superscriptaddress]{revtex4-2} 
\usepackage{array}
\newcolumntype{P}[1]{>{\centering\arraybackslash}p{#1}}
\newcolumntype{N}[1]{>{\centering\arraybackslash}n{#1}}
\usepackage{booktabs}
\usepackage{amsmath}
\usepackage{multirow}
\usepackage{xr-hyper}
\usepackage{hyperref}
\usepackage{csquotes}
\usepackage[dvipsnames]{xcolor}  
\usepackage{soul}
\usepackage{graphicx}
\usepackage{parskip}
\usepackage{natbib}
\setcitestyle{numbers,square,super}
\usepackage{xr}
\usepackage{cleveref}
\usepackage{soul}
\makeatletter
\newcommand*{\addFileDependency}[1]{
  \typeout{(#1)}
  \@addtofilelist{#1}
  \IfFileExists{#1}{}{\typeout{No file #1.}}
}
\makeatother

\begin{document}

\title{Determining the Proximity Effect Induced Magnetic Moment in Graphene by Polarized Neutron Reflectivity and X-ray Magnetic Circular Dichroism}


\author{R. O. M. Aboljadayel}
\email{roma2@cam.ac.uk}
\affiliation{Cavendish Laboratory, Physics Department, University of Cambridge, Cambridge CB3 0HE, United Kingdom.} 
\author{C. J. Kinane}
\affiliation{ISIS Facility, STFC Rutherford Appleton Laboratory, Harwell Science and Innovation Campus, Oxon, OX11 0QX, United Kingdom.}
\author{C. A. F. Vaz}
\affiliation{Swiss Light Source, Paul Scherrer Institut, 5232 Villigen PSI, Switzerland}
\author{D. M. Love}
\affiliation{Cavendish Laboratory, Physics Department, University of Cambridge, Cambridge CB3 0HE, United Kingdom.}
\author{R. S. Weatherup}
\affiliation{Department of Engineering, University of Cambridge, Cambridge CB3 0FA, United Kingdom.}
\author{P. \surname{Braeuninger-Weimer}}
\affiliation{Department of Engineering, University of Cambridge, Cambridge CB3 0FA, United Kingdom.}
\author{M.-B. Martin}
\affiliation{Department of Engineering, University of Cambridge, Cambridge CB3 0FA, United Kingdom.}
\author{A. Ionescu}
\affiliation{Cavendish Laboratory, Physics Department, University of Cambridge, Cambridge CB3 0HE, United Kingdom.}
\author{A. J. Caruana}
\affiliation{ISIS Facility, STFC Rutherford Appleton Laboratory, Harwell Science and Innovation Campus, Oxon, OX11 0QX, United Kingdom.}
\author{T. R. Charlton}
\affiliation{ISIS Facility, STFC Rutherford Appleton Laboratory, Harwell Science and Innovation Campus, Oxon, OX11 0QX, United Kingdom.}
\author{J. Llandro}
\affiliation{Cavendish Laboratory, Physics Department, University of Cambridge, Cambridge CB3 0HE, United Kingdom.}
\author{P. M. S. Monteiro}
\affiliation{Cavendish Laboratory, Physics Department, University of Cambridge, Cambridge CB3 0HE, United Kingdom.}
\author{C. H. W. Barnes}
\affiliation{Cavendish Laboratory, Physics Department, University of Cambridge, Cambridge CB3 0HE, United Kingdom.}
\author{S. Hofmann}
\affiliation{Department of Engineering, University of Cambridge, Cambridge CB3 0FA, United Kingdom.}
\author{S. Langridge}
\email{sean.langridge@stfc.ac.uk}
\affiliation{ISIS Facility, STFC Rutherford Appleton Laboratory, Harwell Science and Innovation Campus, Oxon, OX11 0QX, United Kingdom.}


\date{\today}

\begin{abstract}
We report the magnitude of the induced magnetic moment in CVD-grown epitaxial and rotated-domain graphene in proximity with a ferromagnetic Ni film, using polarized neutron reflectivity (PNR) and X-ray magnetic circular dichroism (XMCD). The XMCD spectra at the C \textit{K}-edge confirms the presence of a magnetic signal in the graphene layer and the sum rules give a magnetic moment of up to $\sim\,0.47\,\mu$\textsubscript{B}/C atom induced in the graphene layer. For a more precise estimation, we conducted PNR measurements. The PNR results indicate an induced magnetic moment of $\sim$ 0.53 $\mu$\textsubscript{B}/C atom at 10 K for rotated graphene and $\sim$ 0.38 $\mu$\textsubscript{B}/C atom at 10 K for epitaxial graphene. Additional PNR measurements on graphene grown on a non-magnetic Ni\textsubscript{9}Mo\textsubscript{1} substrate, where no magnetic moment in graphene is measured, suggest that the origin of the induced magnetic moment is due to the opening of the graphene's Dirac cone as a result of the strong C \textit{p$_z$}-3\textit{d} hybridization.
\end{abstract}


\maketitle

\section{Introduction}
Graphene is a promising material for many technological and future spintronic device applications such as spin-filters,\cite{Karpan2007b,Weser2011c,Zhang2015a,Hogl2020,Meng2013,Song2015,Xu2018} spin-valves and spin field-effect transistors due to its excellent transport properties.\cite{Hill2006c,Semenov2007a} Graphene can have an intrinsic charge carrier mobility of more than 200,000 cm${^2}$V$^{-1}$s$^{-1}$ at room temperature (RT),\cite{sze2006physics} and a large spin relaxation time as a result of its long electron mean free path and its negligible spin-orbit and hyperfine couplings.\cite{Weser2011c,K-Pi1} 

Manipulating spins directly in the graphene layer has attracted great attention as it opens new ways for using this 2D material in spintronics applications.\cite{Weser2011c,Zhang2018} This has been realized via various approaches such as through the proximity-induced effect,\cite{Weser2010b,LiLijun1,Wang2015a,Haugen2008b,Weser2011c} chemical doping of the graphene surface \cite{K-Pi1} or through a chemically-induced sublattice.\cite{Dinh1} Here, we report the feasibility of the first method in utilizing the exchange coupling of local moments between graphene and a ferromagnetic (FM) material to induce a magnetic moment in graphene.  

Graphene is a zero-gap semiconductor because the $\pi$ and $\pi^*$ bands meet at the Fermi energy ($E$\textsubscript{F}), at the corner of the graphene's Brillouin zone ($K$ points), i.e. at degenerate points forming the Dirac point ($E$\textsubscript{D}), where the electronic structure of these bands can be described using the tight-binding model.\cite{Dedkov2015a,Heersche2007a} However, the adsorption of graphene on a strongly interacting metal distorts its intrinsic band structure around $E$\textsubscript{D}. This is a result of the overlap of the graphene's valence band with that of the metal substrate due to the breaking of degeneracy around $E$\textsubscript{D} in a partially-filled \textit{d}-metal, as discussed in the \textit{universal model} proposed  by Voloshina and Dedkov.\cite{Voloshina2014} Their model was supported by density functional theory calculations and proven experimentally using angle-resolved photoemission spectroscopy.\cite{Dedkov2015a,Voloshina2014,Tablero2010a,Dahal2014a,Dedkov2010c,Vita2014,Marchenko2015} Furthermore, a small magnetic signal was detected in the X-ray magnetic circular dichroism (XMCD) spectra of the graphene layer in proximity with a FM transition metal (TM) film, suggesting that a magnetic moment is induced in the graphene.\cite{Weser2010b,Weser2011c,Mertins2018,Mendes2019} However, no direct quantitative analysis of the total induced magnetic moment has been reported.  

It is widely accepted that graphene's C atoms are assembled in what is known as the top-\textit{fcc} configuration on top of close-packed (111) surfaces, where the C atoms are placed on top of the atoms in the first and third layers of the TM substrate.\cite{Voloshina2014,Dahal2014a,Dedkov2010c} The strength of the graphene-TM interaction is influenced by the lattice mismatch, the graphene-TM bond length and the position of the \textit{d} orbital of the TM relative to \textit{E}\textsubscript{F}. Therefore, a Ni(111) substrate was used as the FM since it has a small lattice mismatch of -1.2\%, a bond length of $2.03$ \AA\ and their \textit{d} orbitals are positioned $\sim 1.1$ eV below \textit{E}\textsubscript{F} (i.e. forming $\pi-$\textit{d} hybrid states around the $K$ points).\cite{Voloshina2014,Tablero2010a,Dahal2014a,Karpan2008a} Epitaxial and rotated-domain graphene structures were investigated since rotated graphene is expected to interact more weakly with the TM film underneath. This is a result of the loss of epitaxial relationship and a lower charge transfer from the TM due to missing direct Ni\textsubscript{top}$-$C interaction and the smaller region covered by extended graphene layer as a result of the 3$^{\circ}$ rotation between the graphene and Ni .\cite{Patera2013,Dahal2014a,Weatherup2014a,Kozlov2012} Therefore, a smaller magnetic moment is expected to be induced in rotated-domain graphene.

We have studied the structural, magnetic and electronic properties of epitaxial- and rotated-domain graphene grown on Ni films, confirmed the presence of a magnetic moment in graphene by element-specific XMCD and measured the induced magnetic moment at 10 K and 300 K for rotated and epitaxial graphene using polarized neutron reflectivity (PNR).

To our knowledge, our attempt is the first reported approach in using PNR and XMCD sum rules combined to estimate the total induced magnetic moment in graphene. This is due to the thinness of graphene which is close to the resolution of the PNR technique, and the difficulty in processing the XMCD C \textit{K}-edge signal due to the contribution of the carbon contamination in the beamline optics. We attribute the presence of an induced magnetic moment in graphene to the hybridization of the C \textit{p}\textsubscript{z} orbital with the 3\textit{d} bands of the TM, which is supported by additional PNR measurements on graphene grown on a non-magnetic Ni$_9$Mo$_1$(111) substrate, where no magnetic moment is detected in the graphene layer.

\section{Results and Discussion}
\subsection{Raman Spectroscopy Measurements} \label{Ramansection}
In order to evaluate the quality, number of graphene layers, the doping and defect density in the grown graphene samples we used Raman spectroscopy, which is a non-destructive technique known to be particularly sensitive to the structural and electronic properties of graphene.\cite{Wang2008,Childres2013} For these measurements, the graphene layer was first transferred by a chemical etching process from the metallic film onto a Si substrate with a thermally oxidized SiO$_2$ layer similar to that reported in Ref [\!\!\citenum{Reina2008}]. This was done to avoid loss in the resonance conditions due to the strong chemical interaction between the graphene $\pi$ orbital and the \textit{d}-states of Ni and Ni$_9$Mo$_1$ which also alters the graphene’s \textit{p}\textsubscript{z} orbitals (see Experimental section for further details).

\begin{figure}[t]
\centering
\includegraphics[width=0.9\textwidth]{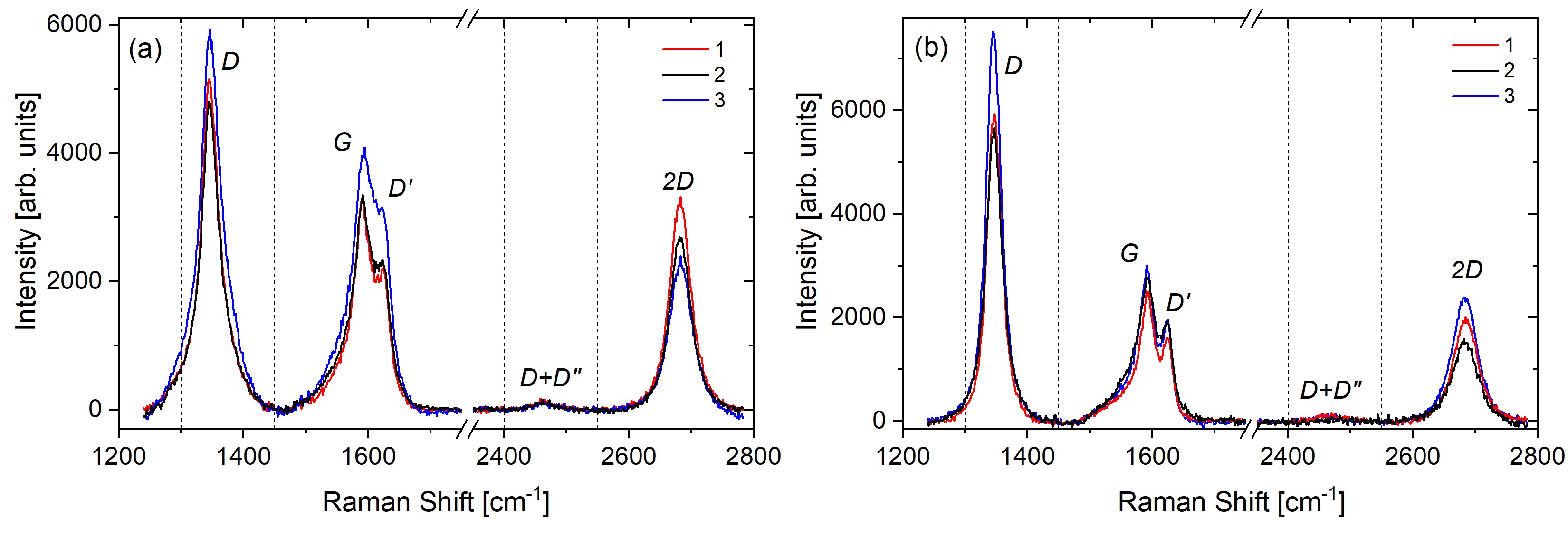}
\caption{Room temperature Raman spectroscopy measurements taken at three different regions (1-3) after transferring the graphene from (a) the epitaxial Ni and (b) rotated Ni samples to a Si/SiO$_2$ wafer, showing the graphene's characteristic peaks. The dashed vertical lines separate the regions of the different peaks.}
\label{RamanSpectra}
\end{figure}

Figure \ref{RamanSpectra} shows the Raman scans taken at three different regions of the graphene after being transferred from the Ni films (see Section \ref{Raman} of the Experimental Section). All the spectra possess the \textit{D}, \textit{G}, \textit{D'}, \textit{D}+\textit{D"} and 2\textit{D} peaks.\cite{Malard2009,Ferrari2013} Although all the 2\textit{D} peaks shown in Figure \ref{RamanSpectra} were fitted with single Lorentzians, they have relatively broad full-width at half-maximum (FWHM). The average FWHM of the 2\textit{D} peak of epitaxial and rotated-domain graphene transferred from the Ni film are 40.8 cm$^{-1}$ and 46.2 cm$^{-1}$, respectively. Furthermore, the spectra of both samples show a high \textit{I}\textsubscript{\textit{D}}/\textit{I}\textsubscript{G} ratio (average of 1.49 for epitaxial graphene and 2.35 for rotated graphene). The variation in the spectra of each sample, presence of second order and defect-induced peaks, the large FWHM of the 2\textit{D} peak and the high \textit{I}\textsubscript{\textit{D}}/\textit{I}\textsubscript{G} ratio could be a result of the chemical etching and transfer process (see the Sample Preparation section) or the chemical doping from the HNO\textsubscript{3} used to etch the metallic films. Therefore, it is difficult to estimate the number of graphene layers based on the position of the \textit{G} and 2\textit{D} peaks and \textit{I$_{2D}$}/\textit{I$_G$} ratio. However, the SEM scan and LEED diffraction pattern in Figure \ref{SEM} (a) and (c), respectively, show a single epitaxial graphene layer grown on Ni(111). The broader FWHM of the 2\textit{D} and the higher average \textit{I$_{2D}$}/\textit{I$_G$} ratio in the rotated graphene compared to the epitaxial structure could be attributed to the formation of more defective or turbostratic (multilayer graphene with relative rotation between the layers) graphene as a result of the occasional overlap of the graphene domains.\cite{Cabrero-Vilatela2016a} The Raman spectra for the graphene/Ni$_9$Mo$_1$ sample, as well as the full list of the peak positions and the 2\textit{D} average FWHM of all the measured samples is provided in the supplementary material (SI).

\subsection{X-ray Magnetic Circular Dichroism (XMCD)}  \label{XMCDsection}
\begin{figure}[t!]
\centering
\includegraphics[width=0.90\textwidth]{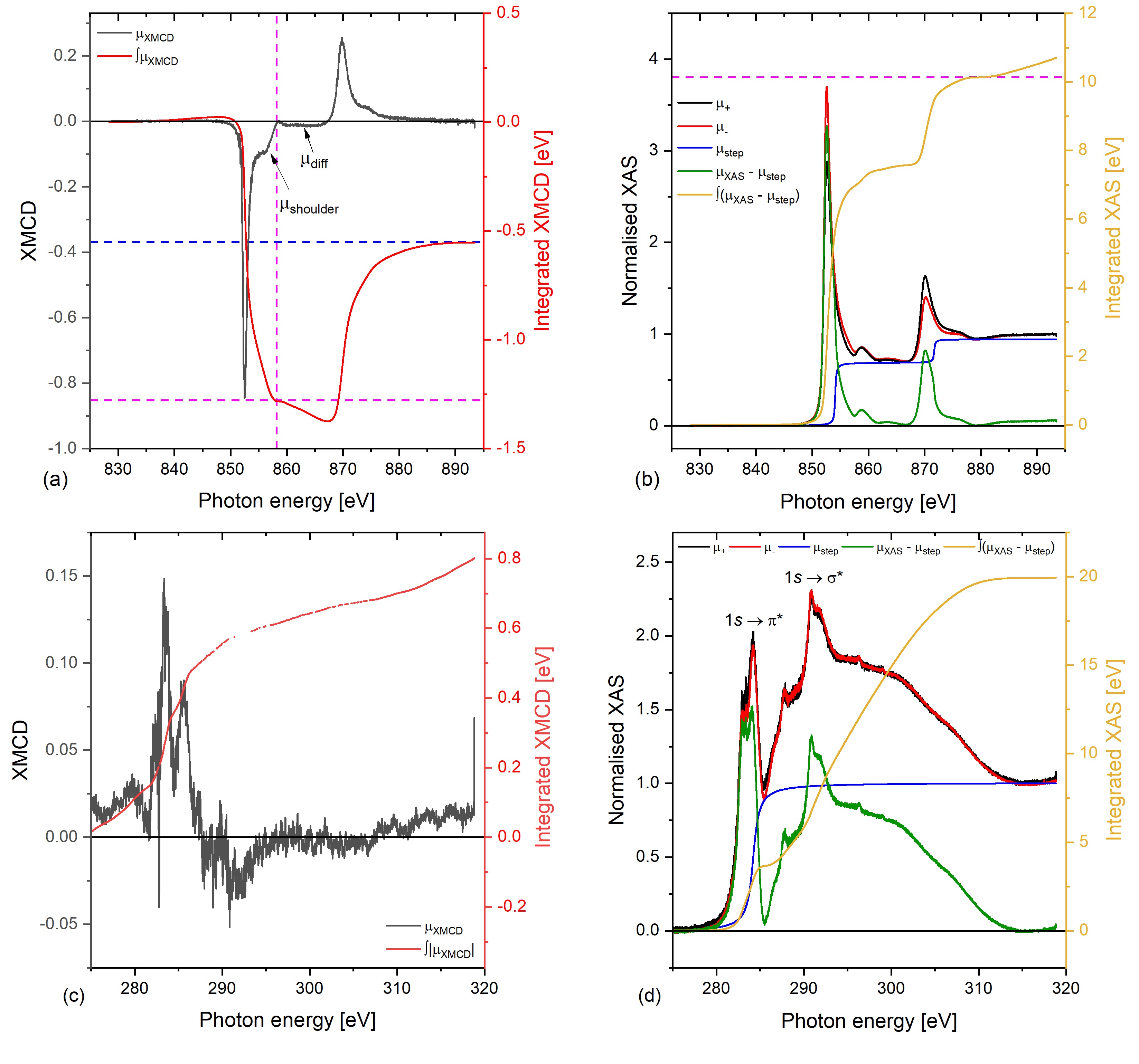}
\caption{X-ray absorption spectra for circular polarized light and the areas used to apply the sum rules for the rotated graphene/Ni sample measured at 300 K: (a) and (b) XMCD and XAS spectra for Ni \textit{L}\textsubscript{2,3}-edge. (c) and (d) XMCD and XAS spectra for the graphene layer. The vertical and horizontal dashed lines in (a) indicate the maximum integration range over the \textit{L}\textsubscript{2,3} peak and the values used for the calculation of $m_o$ and $m_s$, respectively.}
\label{fig:XMCDXASNiCSample}
\end{figure} 
 The X-ray absorption spectra ($\mu$\textsubscript{XAS}) and the XMCD response ($\mu$\textsubscript{XMCD}) at the  Ni \textit{L}\textsubscript{2,3}-edge in the rotated graphene/Ni sample are shown in Figure \ref{fig:XMCDXASNiCSample}. The spectra show no sign of oxidation, proving that graphene acts as a good passivation layer against oxidation.\cite{Martin2015a,Dlubak2012a} The region between the \textit{L}$_3$ and \textit{L}$_2$ edges with a constant negative intensity is known as the diffuse magnetism region, $\mu$\textsubscript{diff}, and it has been observed and reported for Co, Ni and Fe XMCD spectra.\cite{OBrien1994a,Eriksson1992} $\mu$\textsubscript{diff} is expected to arise as a result of the opposite spin directions for the 4\textit{s} and 3\textit{d} electrons, interstitial and \textit{sp}-projected magnetic moments, and the fact that it couples antiferromagnetically to the sample's total magnetic moment in 3\textit{d} elements, except for Mn.\cite{OBrien1994a} Although $\mu$\textsubscript{diff} has been reported to contribute to about $-7\%$  to the total magnetic moment in Ni,\cite{OBrien1994,OBrien1994a} since the sum rule does not account for $\mu$\textsubscript{diff}, the integration range over the \textit{L}$_3$ was stopped just before $\mu$\textsubscript{diff} for the calculation of the orbital magnetic moment, $m_o$, and the spin magnetic moment, $m_s$ (858.7 eV). On the other hand, the main \textit{L}$_3$ peak and the shoulder, $\mu$\textsubscript{shoulder}, are due to multiple initial-states configuration, $3d^8$ and $3d^9$, respectively, and therefore they were accounted for in the sum rule calculations.\cite{OBrien1994a} 
 
 Although the non-resonant contribution was subtracted from the $\mu_+$ and $\mu_-$ spectra, a higher background is measured at the post-edge ($E> 880$ eV). This tail has been excluded from the sum rules as it is considered part of the non-resonant contribution. The calculated $m_o$ and $m_s$ are 0.084 $\mu$\textsubscript{B}/Ni atom and 0.625 $\mu$\textsubscript{B}/Ni atom, respectively, and thus $m_{total}$ is 0.709 $\mu$\textsubscript{B}/Ni atom (see the SI for the expressions for $m_o$, $m_s$ and $m_{total}$ at the \textit{L}-edge). Considering the 20$\%$ accuracy of the XMCD technique in estimating the magnetic moments of materials, the results obtained for Ni is consistent with the values reported in the literature. \cite{Eriksson90,Vaz2008d} 

The C \textit{K}-edge spectra for the rotated graphene/Ni sample are shown in Figure \ref{fig:XMCDXASNiCSample} (c) and (d). For the C \textit{K}-edge, higher sources of errors are expected in the XMCD estimation attributed to the difficulty of applying the sum rules to the C \textit{K}-edge spectra in comparison with that for Ni \textit{L}\textsubscript{2,3}-edge. For instance, various studies have been reported for Ni \cite{Weser2010b,OBrien1994,Nakajima1999a,Dedkov2010c,Eriksson90,Chen1990a,Soderlind1992} which can be used as references for our measurements, but the application of the sum rules has not been reported for graphene before. Also, the number of holes, $n_h$, has not been measured for C previously. Moreover, the gyromagnetic factor ($g$) of the graphene was found to be different depending on the underlying substrate,\cite{Menezes2017,Song2010,Semenikhin2020} and it has not been reported for graphene on Ni. It is also noteworthy to mention the difficulty associated with measuring the C \textit{K}-edge due to the C contamination of the optical elements which appear as a significant reduction in the incoming intensity at this particular energy.   

Nonetheless, we can obtain an upper limit to the orbital moment of the graphene layer by integrating the modulus of the dichroic signal, $|\mu_{XMCD}|$, which is shown in Figure \ref{fig:XMCDXASNiCSample} (c), red curve. Although the magnetic dichroism response is expected mainly at the peak corresponding to the 1\textit{s}$\rightarrow \pi^*$ transition as a result of the C \textit{p}\textsubscript{z}$-$Ni 3\textit{d} hybridization,\cite{Weser2010b} a small magnetic signal is observed at the 1\textit{s}$\rightarrow \sigma^*$ transition peak as well; a similar behaviour was reported for graphene/2 ML Co/Ir(111).\cite{Vita2014} The  calculated upper bound $m_o$ for graphene is 0.062 $\mu$\textsubscript{B}/C atom using $n_h=4$; which corresponds to an $m_s$ of 0.412 $\mu_B$/C atom, using $g=2.3$, which is the value reported for graphene grown on SiC.\cite{Menezes2017} Therefore, $m_{total}$ of the rotated-domain graphene grown on Ni is $\sim$0.474 $\mu_B$/C atom (see the SI for the expressions of $m_o$, $m_s$ and $m_{total}$ at the \textit{K}-edge). 

Despite the large uncertainties expected for the estimated graphene moments, the XMCD results demonstrate the presence of magnetic polarization in graphene. For more precise, quantified and independent estimates, we turn to PNR.

%
\subsection{Polarized Neutron Reflectivity} \label{PNRsection}
PNR experiments were carried out to measure the magnetic properties of each layer of the samples individually and to determine the value of the induced magnetic moment in graphene quantitatively.

\begin{figure}[b!]
\centering
\includegraphics[width=1.0\textwidth]{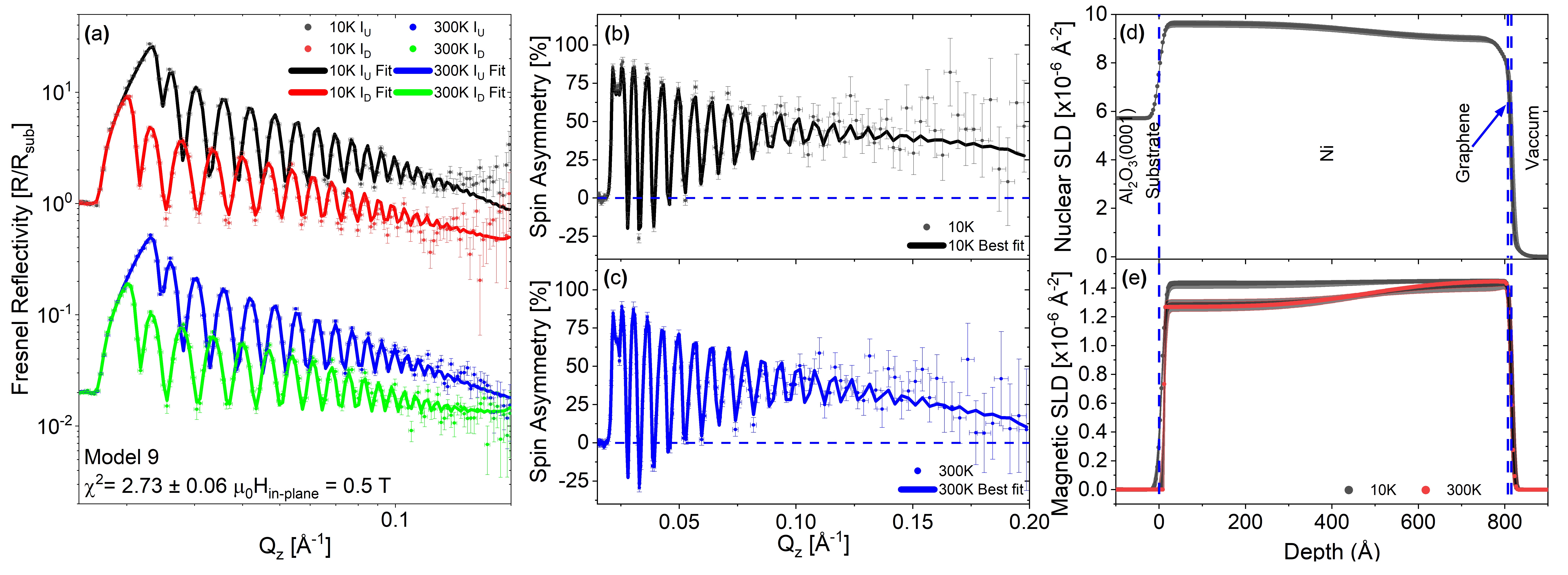}
\caption{PNR Model 9: consists of a Ni layer split into two regions for the rotated graphene/Ni sample with a magnetic dead-layer at the substrate interface and continuous magnetic moment variation up to the graphene layer which is allowed to be magnetic. (a) Fresnel reflectivity for 10 K and 300 K, (b) and (c) show the spin asymmetries, (d) is the nuclear scattering length density (nSLD) profile and (e) the magnetic scattering length density (mSLD) profile. The grey banded regions around the SLD lines are the 95\% Bayesian confidence intervals.}
\label{RotGrNiPNR}
\end{figure}

\begin{figure}[t!]
\centering
\includegraphics[width=1.0\textwidth]{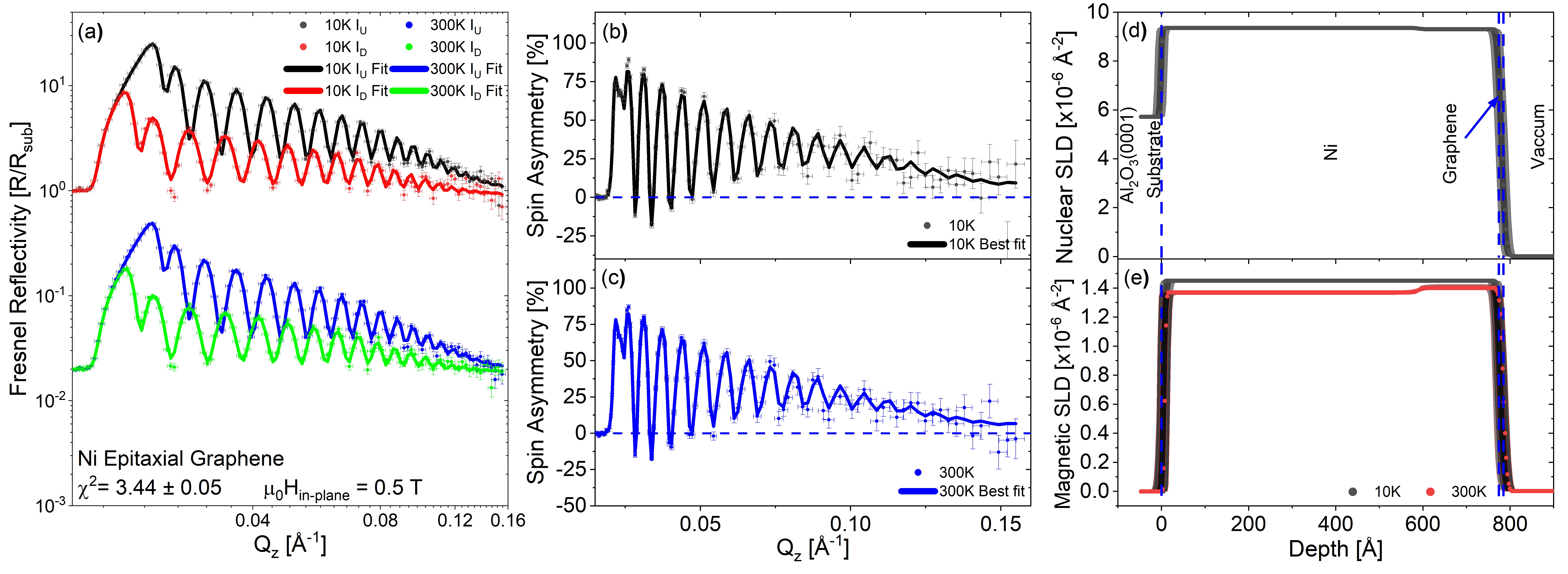}
\caption{Ni layer split into two regions for the epitaxial graphene/Ni sample (a) Fresnel reflectivity for 10 K and 300 K, (b) and (c) show the spin asymmetries, (d) is the nuclear scattering length density (nSLD) profile and (e) the magnetic scattering length density (mSLD) profile.}
\label{EpiGrNiPNR}
\end{figure}

The PNR results for the rotated-domain graphene/Ni(111) and epitaxial graphene/Ni(111) samples, measured at 10 K and 300 K, are displayed in Figures \ref{RotGrNiPNR} and \ref{EpiGrNiPNR}, respectively. Panel (a) for each figure shows the Fresnel reflectivity profiles and panels (b) and (c) the spin asymmetry ($SA=[R_+-R_-]/[R_++R_-]$, where $R_+$ and $R_-$ are the spin-up and spin-down neutron specular reflectivities, respectively). $SA$ scales with the magnetic signal. A flat $SA$ line at zero, shown as a blue dashed line, represents no net magnetic induction present in the system. Panel (d) displays the nuclear scattering length density (nSLD), for the sample structure. This structure is shared at both temperatures in the co-refinement, and panel (e) shows the magnetic scattering length density (mSLD) for each temperature.

The fitting procedure is fully described in the SI, which contains the various models used to fit the rotated graphene/Ni sample. The 10 K and 300 K data shown in Figures \ref{RotGrNiPNR} and \ref{EpiGrNiPNR} were fitted simultaneously with a shared nSLD and independent mSLD. The bulk of the fitting model selection was done on the rotated graphene/Ni sample. The study presented in the SI show the importance of prior knowledge of the sample properties to obtain the best PNR fit to estimate the induced magnetic moment in graphene. We used the information obtained from the structural characterizations (SEM and Raman spectroscopy) to obtain a lower bound on the graphene layer thickness. The fit tends to shrink the graphene thickness to less than one monolayer, if unrestrained, which does not agree with the SEM and Raman scans (refer to SI). This could be due to the limited $Q$ (wavevector transfer) range measured in the time available for the experiment, as the fit is found to rely strongly on the high $Q$ statistics. It should be noted that in model 9 shown in Figure \ref{RotGrNiPNR} the magnetic moment of the graphene layer was allowed to fit to zero, but the analysis always required a non-zero value for the magnetic moment to obtain a good fit with low uncertainty, which is consistent with the XMCD results. The unrestrained graphene thickness models are shown in the SI. It is noteworthy that the improvement in the figure of merit ($\chi^2$) for the unconstrained to the constrained graphene thickness are within the error bar of each other. However, the results show that the graphene thickness has only a subtle influence, if any, on the amount of the induced magnetization as can be deduced from the values of the measured magnetic moments. To get greater certainty on the model selection given the data we used a nested sampler\cite{doi:10.1063/1.1835238,McCluskey_2020} to calculate the Bayesian evidence taking into account the structural information, providing a high degree of confidence in the final fit model as shown in the SI.

Additional scenarios were also tested. For example, oxidation of the Ni layer and the formation of Ni-carbide were examined by embedding an intermediate NiO and Ni$_2$C layer, respectively, at the interface between the Ni and graphene, but this lead to poorer fits and the best results, shown in Figures \ref{RotGrNiPNR} and \ref{EpiGrNiPNR}, were achieved using a simpler model: Substrate/a FM layer split into two regions/graphene (see the modelling methodology in the SI).

The results of the fits to the PNR data are summarised in Table \ref{PNRSummary}. Interestingly, it is the rotated-domain graphene rather than the epitaxially grown graphene that has the largest induced magnetic moment, counter to the initial hypothesis that the coupling between the graphene and Ni(111) would be weaker in the rotated-domain case. The greater uncertainty in the values of the moment obtained for the graphene in the epitaxial case can be attributed to the Bayesian analysis being sensitive to the reduced count time and short $Q$ range that the epitaxial Ni was measured with, due to finite available beam time. Both Ni(111) samples have the full Ni moment at 10 K, which is only slightly reduced at 300 K. This reduction is associated with a magnetic gradient across the Ni(111) film. In the rotated-domain sample, this gradient, as shown in Figure \ref{RotGrNiPNR} (d), starts with a magnetic moment close to the bulk nSLD of Ni (9.414 x 10$^{-6}\,$\AA$^{-2}$) and reduces in size towards the surface, and is required in order to fit the lower $Q$ features and allow the higher $Q$ features to converge, paramount to getting certainty on the thin graphene layers. The mSLD consequently also has a gradient that oppositely mirrors the nSLD at 300 K and reaches the full moment for Ni (0.6 $\mu_{B}$/Ni atom equivalent to an mSLD = 1.4514 x 10$^{-6}\,$\AA$^{-2}$) near the surface and being slightly reduced near the substrate. The Ni(111) used for the epitaxial graphene, displays a much weaker gradient, being almost uniform across the Ni thickness but has the same general trends. We attribute the origin of the difference in the nSLD profiles to the fact that the samples were deposited at different times, following the growth recipe described in the Sample Preparation section.

The Ni$_{9}$Mo$_{1}$ sample was used to clarify whether the induced magnetic moments in graphene are due to the C \textit{p}\textsubscript{z}-3\textit{d} hybridization which result in opening the Dirac cone, as postulated in Refs. [\!\!\citenum{Dedkov2010c,Weser2011c,Karpan2008a,Dedkov2015a,Voloshina2014}], or because of electron transfer (spin doping) and surface reconstruction, which distorts the \textit{d} band of the TM as for fullerene/non-magnetic TM, as proposed in Refs. [\!\!\citenum{MaMari2015a,Pai2010,Moorsom2014}]. For this purpose, a Ni$_9$Mo$_1$ film was used with the aim of preserving the \textit{fcc} crystal structure of Ni while suppressing its magnetization, as suggested in Ref. [\!\!\citenum{Hansen1958}]. Since the Ni is doped by 10$\%$ only, the lattice mismatch and bond length to graphene are expected to be similar to that for graphene/Ni(111) sample, but the \textit{d}-orbital position is considerably downshifted with respect to \textit{E}\textsubscript{F}. The growth procedure and the structural properties of the sample is discussed in detail in the Experimental Section.

\begin{figure}[t!]
\centering
\includegraphics[width=1.0\textwidth]{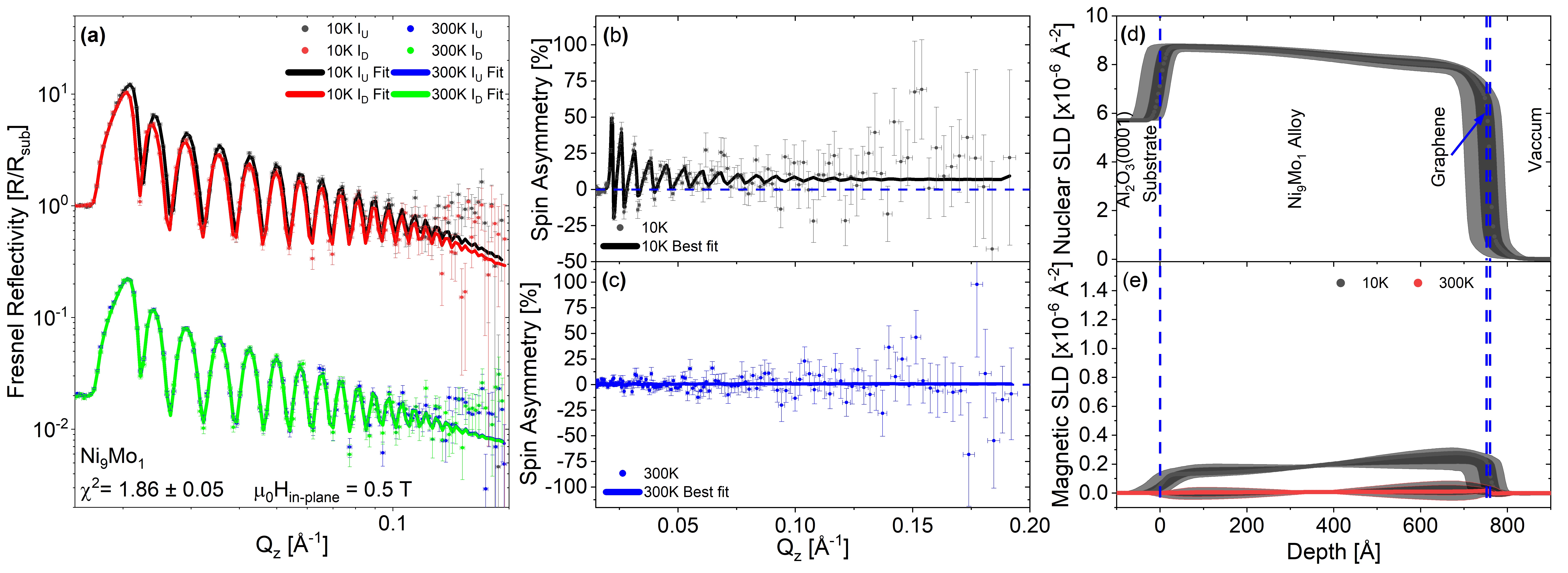}
\caption{Ni$_{9}$Mo$_{1}$ split into two regions for the graphene/Ni$_9$Mo$_1$ sample. (a) Fresnel reflectivity for 10 K and 300 K,  (b) and (c) show the spin asymmetries, (d) is the nuclear scattering length density (nSLD) profile and (e) the magnetic scattering length density (mSLD) profile.}
\label{RotGrNiMo}
\end{figure}

\begin{table}[b!]
\caption{Summary of the PNR results for the rotated graphene/Ni, epitaxial graphene/Ni and graphene/Ni$_9$Mo$_1$ samples using model 9: sapphire/a FM layer split into two regions/graphene. The values in the parenthesis are the lower and upper 95\% Bayesian confidence limits.\cite{Bumps_URL_ref}}
\centering
\begin{tabular}{p{2.2cm} p{2.0cm} p{2.5cm} p{3.6cm} p{2.7cm} c}
\hline
 & \multicolumn{3}{c}{FM Layer1 + FM layer2} & \multicolumn{2}{c}{Graphene}\\	
 \hline
 \centering{Sample} & \centering{Temperature} & \centering{Thickness} & \centering{Magnetic Moment} & \centering{Thickness} & Magnetic Moment\\
 & \centering{[K]} & \centering{[nm]} & \centering{[$\mu_B$/atom]} & \centering{[nm]} & [$\mu_B$/atom]\\
\hline
\centering{\multirow{2}{*}{Rotated Gr/Ni}} & \centering{10 K} &\centering{\multirow{2}{*}{$80.6(80.2, 81.3)$}} & \centering{$0.60(0.60, 0.61)$} & \centering{\multirow{2}{*}{$0.86(0.81, 1.04)$}} & $0.53(0.52, 0.54)$\\
& \centering{300 K} & & \centering{$0.57(0.56, 0.57)$} & & $0.53(0.51, 0.54)$ \\
\hline
\centering{\multirow{2}{*}{Epitaxial Gr/Ni}} & \centering{10 K} &\centering{\multirow{2}{*}{$77.5(77.1, 78.0)$}} & \centering{$0.61(0.60, 0.61)$} & \centering{\multirow{2}{*}{$0.92(0.2, 1.2)$}} & $0.38(0.12, 0.50)$ \\
 & \centering{300 K} & & \centering{$0.579(0.575, 0.583)$} & & $0.2(0.0, 0.5)$ \\
 \hline
\centering{\multirow{2}{*}{Gr/Ni$_{9}$Mo$_{1}$}} & \centering{10 K} & \centering{\multirow{2}{*}{$75.0(74.0, 76.4)$}} & \centering{$0.064(0.056, 0.071)$} & \centering{\multirow{2}{*}{$0.84(0.80, 1.00)$}} & $0.12(0.07, 0.17)$ \\
 & \centering{300 K} & & \centering{$0.003(-0.004, 0.01)$} & & $0.02(0.001, 0.05)$ \\
\hline
\end{tabular}
\label{PNRSummary}
\end{table} 

The results of the PNR measurements of the graphene/Ni$_9$Mo$_1$ sample are shown in Figure \ref{RotGrNiMo}. Again there is a gradient across the Ni$_9$Mo$_1$ film. At 10 K a small, but detectable spin splitting and a minute variation in the $SA$ are observed. Surprisingly, a higher magnetization is detected in graphene ($0.12\,(0.07, 0.17$) $\mu$\textsubscript{B}/C atom) than in the Ni$_9$Mo$_1$ film ($0.064\,(0.06, 0.07$) $\mu$\textsubscript{B}/C atom), but the 95\% confidence intervals indicate that we are not sensitive to this to a degree to say for sure given the \textit{Q} range and counting statistics in the data. All we can ascertain is that there is a small moment in the Ni$_9$Mo$_1$ and a non zero moment in the graphene. At 300 K, both the alloy and the graphene have effectively zero moment within the 95\% confidence intervals. The origin of the small residual moment at 10 K could arise from clusters of unalloyed Ni throughout the layer that become ferromagnetic at low temperature, which then polarize the graphene as per the Ni(111) samples. At 300 K, no magnetic moment in the Ni$_9$Mo$_1$ and graphene are detected. Therefore, the results support the hypothesis of the \textit{universal model}, whereby the measured induced magnetic moment in graphene is due to the opening of the \textit{E}\textsubscript{D} rather than the distortion of the \textit{d} band. This is because no magnetic moment is detected in the Ni$_9$Mo$_1$ nor in the graphene layer.

The PNR fits have shown that at 10 K a magnetic moment of $\sim 0.53\,\mu$\textsubscript{B}/C atom ($\sim 0.38\,\mu$\textsubscript{B}/C atom) was induced in the rotated-domain (epitaxial) graphene grown on Ni films. These results indicate larger moments than previously reported.\cite{Mertins2018,Weser2010b} In Ref. [\!\!\citenum{Mertins2018}], Mertins \textit{et al.} estimated the magnetic moment of graphene to be $0.14\pm0.3\,\mu$\textsubscript{B}/C atom when its grown on a \textit{hcp} Co(0001) film. They obtained this value by comparing the XMCD reflectivity signal of graphene to that of the underlayer Co film.\cite{Mertins2018} Weser \textit{et al.} suggested a similar value ($0.05-0.1 \,\mu$\textsubscript{B}/C atom) for the magnetic moment induced in a monolayer of graphene grown on a Ni(111) film.\cite{Weser2010b} However, their assumption was based on comparing the graphene/Ni system with other C/3\textit{d} TM structures, such as a C/Fe multilayer with 0.55 nm of C,\cite{Mertins2004a} and carbon nanotubes (CNTs) on a Co film,\cite{Cespedes2004a} where a magnetic moment of 0.05 $\mu$\textsubscript{B}/C atom and 0.1 $\mu$\textsubscript{B}/C atom was estimated, respectively. Nonetheless, it is difficult to compare graphene-based heterostructures with other C allotropes/TM systems. This is because the Dirac cone is a characteristic feature of graphene and CNTs of the carbon allotropes. Therefore, one cannot exclude that a different mechanism other than the break of degeneracy around \textit{E}\textsubscript{D} may be responsible for the magnetic moment detected in the C layer of a C/Fe multilayer system. On the other hand, for the CNTs/Co heterostructure, although Dirac cones exist in CNTs, a direct quantitative analysis of the induced magnetic moment was not possible from the MFM images reported in Ref. [\!\!\citenum{Cespedes2004a}]. In contrast, PNR provides a direct estimation of the induced magnetic moment in graphene.




\section{Conclusion}
In summary, we have successfully grown graphene by chemical vapor deposition (CVD) on different TM substrates. Induced magnetic moment in rotated-domain graphene as a result of the proximity effect in the vicinity of a FM substrate was detected by element-specific XMCD measurements at the C \textit{K}-edge. PNR experiments were carried out to determine the magnitude of the magnetic moment detected by XMCD. Although a higher magnetic moment was expected to be induced in epitaxial graphene/Ni sample, the PNR results indicate the epitaxial graphene film had a magnetic moment of $\sim 0.38\,\mu$\textsubscript{B}/C atom as compared to the rotated-domain graphene $\sim 0.53\,\mu$\textsubscript{B}/C atom. Both values are higher than those predicted in other studies.\cite{Weser2010b,Mertins2004a,Cespedes2004a} PNR measurements on graphene/Ni$_9$Mo$_1$ support the \textit{universal model} proposed by Voloshina and Dedkov that the induced magnetic moment in graphene arises as a result of the opening of the graphene's Dirac cone as a result of the strong C \textit{p}\textsubscript{z}-Ni 3\textit{d} hybridization. Our PNR provides the first quantitative estimation of the induced magnetization in graphene by PNR.

\section{Experimental Section}
\subsection{Sample Preparation}
The sample preparation procedure involved two stages; the growth of the TM films using magnetron sputtering and the growth of graphene by CVD. 

\begin{center}
\begin{figure}[b!]
\includegraphics[width=0.80\textwidth]{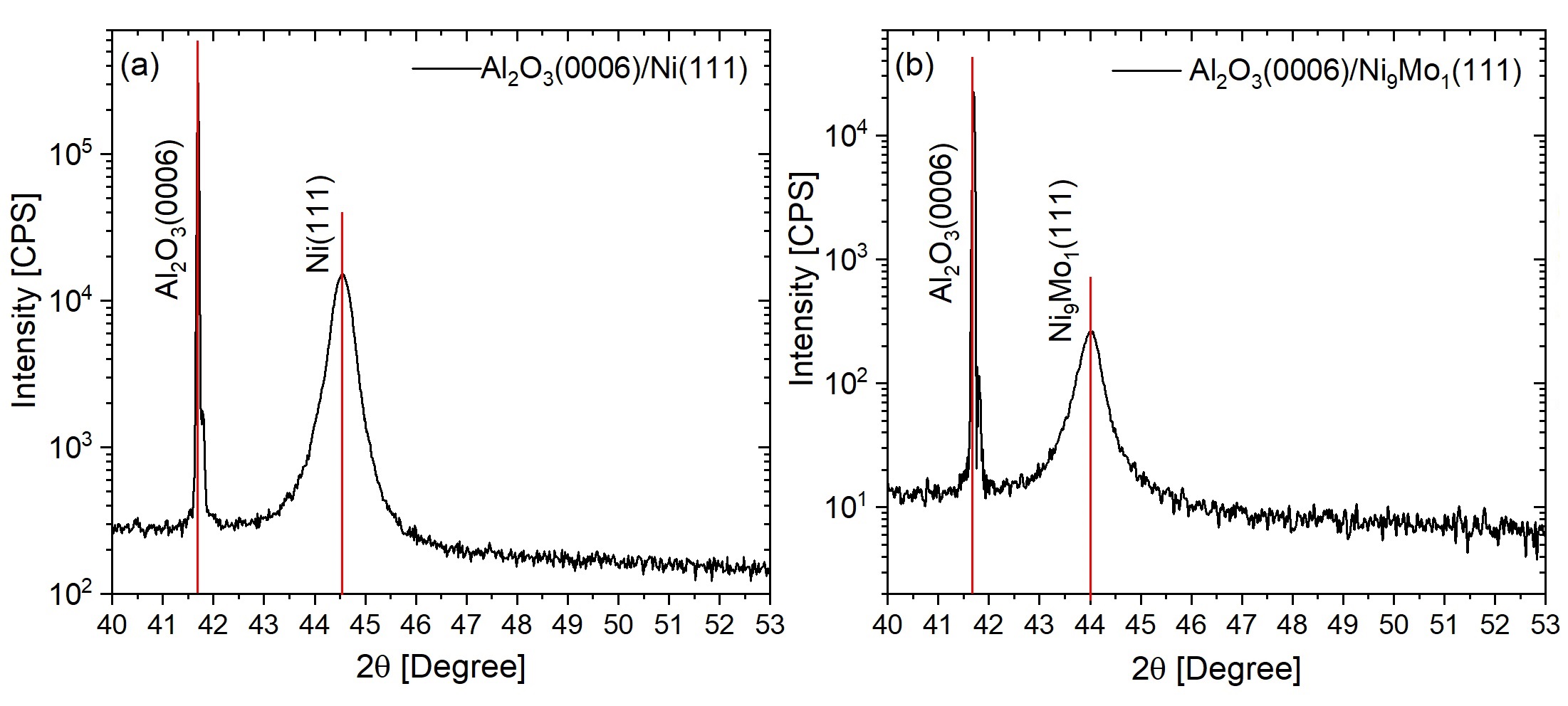}
\caption{X-ray diffraction measurements (scanning range of 40$^{\circ}$ - 53$^{\circ}$) of highly textured films: (a) Al$_2$O$_3$(0001)/Ni(111) and (b) Al$_2$O$_3$(0001)/Ni$_9$Mo$_1$(111).}
\label{XRD}
\end{figure}
\end{center}

The TM films were deposited at RT on 1 mm thick Al$_2$O$_3$(0001) substrates using a CEVP magnetron sputtering chamber with a base pressure of 1.2 - 2 $\times$ 10$^{-8}$ mTorr. The thick substrates were used to reduce the possibility of sample deformation which could affect the reflectivity measurements. The deposition of the TM films was performed using 99.9\% pure Ni and Ni$_9$Mo$_1$ targets. A DC current of 0.1 A and a constant flow of pure Argon of 14 sccm were used to grow 80 nm of highly textured Ni(111) and Ni$_9$Mo$_1$(111) films at a rate of 0.02 nm s$^{-1}$ in a plasma pressure of 2 mTorr (3 mTorr for Ni$_9$Mo$_1$). Figure \ref{XRD} shows the X-ray diffraction (XRD) measurements of the deposited films acquired with a Bruker D8 Discover HRXRD with a Cu K$\alpha$ monochromatic beam (40 kV, 40 mA). The scans show highly textured pure films oriented in the [111] direction for Ni and Ni$_9$Mo$_1$ films.

The samples were then transferred into a CVD system for the growth of graphene directly on Ni(111) and Ni$_9$Mo$_1$(111) films on $\sim$ 2 cm $\times$ 2 cm substrates. Growth recipes similar to those reported by Patera \textit{et al}. \cite{Patera2013} were adapted to obtain epitaxial and rotated-domain graphene directly on the Ni film. For the Ni$_9$Mo$_1$ film, rotated-domain graphene was grown by first introducing pure H$_2$ gas at a rate of 200 sccm to the CVD chamber with a base pressure of 2.7 $\times$ 10$^{-6}$ mbar. Then, the CVD growth chamber was heated to 650 $^{\circ}$C for 12 minutes and the sample was then exposed to C$_2$H$_4$ with a flow rate of 0.24 sccm and 40 minutes before it was cooled down to RT in vacuum. This approach reduces any oxidized TM back to a clean metallic surface before the growth of graphene.
\begin{center}
\begin{figure}[t!]
\includegraphics[width=1.0\textwidth]{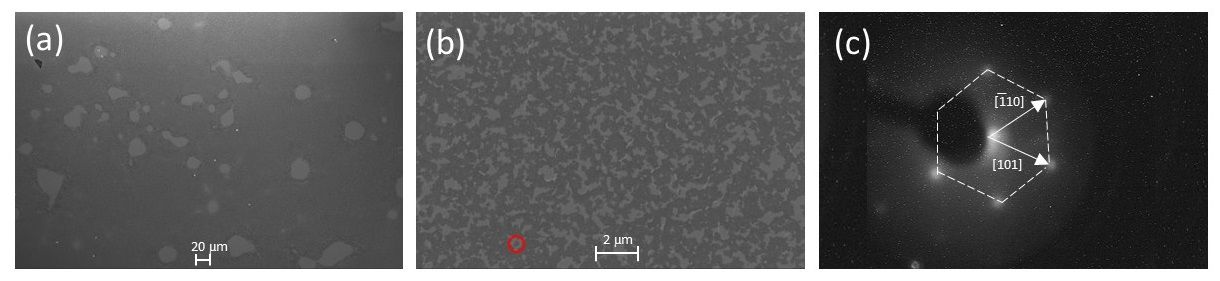}
\caption{SEM images at 1 kV showing the graphene domains for (a) epitaxial graphene/Ni and (b) rotated graphene/Ni and (c) The LEED diffraction pattern of epitaxial graphene on a Ni(111) substrate at 300 eV. The red circle in (b) highlights a single graphene domain with a diameter of $\sim 0.25\,\mu$m.}
\label{SEM}
\end{figure}    
\end{center}

The SEM images shown in Figure \ref{SEM} illustrate the structure of epitaxial and rotated graphene on Ni with a surface coverage of $\sim 70\%-90$\%. Figure \ref{SEM} (a) shows a homogenous monolayer of graphene on Ni. The darker grey regions in Figure \ref{SEM} (b) are the differently oriented graphene domains, whereas the bright areas in Figure \ref{SEM} (a) and (b) are the bare Ni film. The low-energy electron diffraction (LEED) in Figure \ref{SEM} (c) shows the graphene's hexagonal pattern epitaxially grown on the Ni(111) substrate. The ($1\times 1$) grown graphene structure is confirmed since no additional diffraction spots are observed in the LEED pattern.

\subsection{Raman Spectroscopy Measurements} \label{Raman}
RT Raman scans were taken at three different regions of each sample using a 532 nm excitation laser wavelength ($50 \times$ objective lens and $\sim\,1\,\mu$m laser spot size). Before the measurements, the graphene was first transferred by a chemical etching process from the metallic films onto a Si substrate with a 300 nm thermally oxidized SiO$_2$ layer. This approach overcome the fact that closely lattice-matched films lead to a loss in the resonance conditions for observing Raman spectra as a result of the strong chemical interaction between the graphene $\pi$ orbital and the \textit{d}-states of Ni and Ni$_9$Mo$_1$ which also alters the graphene's \textit{p\textsubscript{z}} orbitals. Furthermore, the increase in the C$-$C bond length to match the lattice of the FM leads to significant changes in the graphene's phonon spectrum.\cite{Dahal2014a,Allard2010} For the transfer process, the samples were cleaved into $\sim$ 5 mm $\times$ 5 mm squares and HNO$_3$, diluted to 5\% for Ni$_9$Mo$_1$ and 10\% for Ni, was used to etch the metallic films slowly while preserving the graphene layer.

\subsection{X-ray Magnetic Circular Dichroism (XMCD)}
We carried out element selective XMCD measurements to detect and distinguish the magnetization in the graphene from that of the FM layer. The XMCD experiments were performed at 300 K at the SIM end station of the Swiss Light Source (SLS) at the Paul Scherrer Institut (PSI), Switzerland, using total electron yield (TEY) detection mode with 100\% circularly polarized light. The rotated-domain graphene/Ni sample was set to an incident angle of 30$^{\circ}$ from the incoming X-ray beam. An electromagnet was fixed at 40$^{\circ}$ to the incoming X-ray beam and a magnetic field of 0.11 T was applied for 30 seconds in-plane to the surface of the samples to align the film magnetization along the beam direction. It was then reduced to 0.085 T during the X-ray absorption spectroscopy measurements. The intensity of the incident X-ray beam was measured with a clean, carbon free, gold mesh placed just before the sample position. This is particularly important for normalizing the signal at the C \textit{K}-edge due to the presence of carbon on the surface of the X-ray optical components. 

\subsection{Polarized Neutron Reflectivity}
The PNR measurements were conducted at 10 K and 300 K, under an in-plane magnetic field of 0.5 T, using the Polref instrument at ISIS spallation neutron source (UK). The fitting of the data was done using the \textit{Refl1D} \cite{Refl1d_URL_ref} software package with preliminary fits done in \textit{GenX}.\cite{Bjorck2007} Although Ni is ferromagnetic at RT, the 10 K measurements are expected to provide a better estimation of the induced magnetic moment due to the lower thermal excitations of the electron spin at low temperature. Both the 10 K and 300 K data sets were fitted simultaneously to provide further constraint to the fits. This is analogous to the isotropic contrast matching \cite{Majkrzak2003,KoutsioubasJAC2019} used in soft matter neutron reflectivity experiments. This is very important in this case due to the attempt to measure a thin layer of graphene within a limited total \textit{Q} range for PNR. The PNR is sensitive to only part of the broad fringe from the graphene layer which acts as an envelope function on the higher frequency fringes from the thicker Ni layer underneath (the modelling methodology is discussed in detail in the SI). 

\medskip

\medskip
\textbf{Acknowledgements} \par 
We would like to thank the ISIS Neutron and Muon Source for the provision of beam time (RB1510330 and RB1610424). The data is available at the following DOIs \url{https://doi.org/10.5286/ISIS.E.RB1510330} and \url{https://doi.org/10.5286/ISIS.E.RB1610424}. Part of this work was performed at the Surface and Interface Microscopy (SIM) beamline of the Swiss Light Source (SLS), Paul Scherrer Institut (PSI), Villigen, Switzerland.

Other data presented in this study are available from the corresponding author upon request.

\bibliography{library.bib}
\end{document}


\title{\centering
	Supplementary Material\\ \vspace{0.5cm}
for \\ \vspace{0.5cm}
Determining the Proximity Effect Induced Magnetic Moment in Graphene by Polarized Neutron Reflectivity and X-ray Magnetic Circular Dichroism
\vspace{0.5cm}}

\author{R. O. M. Aboljadayel}
\email{roma2@cam.ac.uk}
\affiliation{Cavendish Laboratory, Physics Department, University of Cambridge, Cambridge CB3 0HE, United Kingdom.} 
\author{C. J. Kinane}
\affiliation{ISIS Facility, STFC Rutherford Appleton Laboratory, Harwell Science and Innovation Campus, Oxon, OX11 0QX, United Kingdom.}
\author{C. A. F. Vaz}
\affiliation{Swiss Light Source, Paul Scherrer Institut, 5232 Villigen PSI, Switzerland}
\author{D. M. Love}
\affiliation{Cavendish Laboratory, Physics Department, University of Cambridge, Cambridge CB3 0HE, United Kingdom.}
\author{R. S. Weatherup}
\affiliation{Department of Engineering, University of Cambridge, Cambridge CB3 0FA, United Kingdom.}
\author{P. \surname{Braeuninger-Weimer}}
\affiliation{Department of Engineering, University of Cambridge, Cambridge CB3 0FA, United Kingdom.}
\author{M.-B. Martin}
\affiliation{Department of Engineering, University of Cambridge, Cambridge CB3 0FA, United Kingdom.}
\author{A. Ionescu}
\affiliation{Cavendish Laboratory, Physics Department, University of Cambridge, Cambridge CB3 0HE, United Kingdom.}
\author{A. J. Caruana}
\affiliation{ISIS Facility, STFC Rutherford Appleton Laboratory, Harwell Science and Innovation Campus, Oxon, OX11 0QX, United Kingdom.}
\author{T. R. Charlton}
\affiliation{ISIS Facility, STFC Rutherford Appleton Laboratory, Harwell Science and Innovation Campus, Oxon, OX11 0QX, United Kingdom.}
\author{J. Llandro}
\affiliation{Cavendish Laboratory, Physics Department, University of Cambridge, Cambridge CB3 0HE, United Kingdom.}
\author{P. M. S. Monteiro}
\affiliation{Cavendish Laboratory, Physics Department, University of Cambridge, Cambridge CB3 0HE, United Kingdom.}
\author{C. H. W. Barnes}
\affiliation{Cavendish Laboratory, Physics Department, University of Cambridge, Cambridge CB3 0HE, United Kingdom.}
\author{S. Hofmann}
\affiliation{Department of Engineering, University of Cambridge, Cambridge CB3 0FA, United Kingdom.}
\author{S. Langridge}
\email{sean.langridge@stfc.ac.uk}
\affiliation{ISIS Facility, STFC Rutherford Appleton Laboratory, Harwell Science and Innovation Campus, Oxon, OX11 0QX, United Kingdom.}

\maketitle

\section{Raman Measurements for the Graphene/N\lowercase{i}$_9$M\lowercase{o}$_1$ sample}

Graphene has two main characteristic peaks in the Raman spectra, a first-order Raman scattering (RS) \textit{G} peak at $\sim$ 1582 cm$^{-1}$, which is a graphite-like line that can be observed in different carbon-based materials;\cite{Shlimak2015a,Allard2010,Malard2009} and a second-order (double-resonance) RS 2\textit{D} peak at $\sim$ 2700 cm$^{-1}$, which is usually used as an indication of a perfect crystalline honeycomb-like structure.\cite{Shlimak2015a,Malard2009} Graphene also possesses other second-order RS peaks such as \textit{D+D"} located at $\sim\,2450$ cm$^{-1}$ and 2\textit{D'} at $\sim$ 3200 cm$^{-1}$ and disorder-induced peaks such as \textit{D} at $\sim$ 1350 cm$^{-1}$ and \textit{D'} at $\sim\,1600$ cm$^{-1}$.\cite{Malard2009,Ferrari2013}

\begin{figure*} [t!]
\centering    
\includegraphics[width=0.5\textwidth]{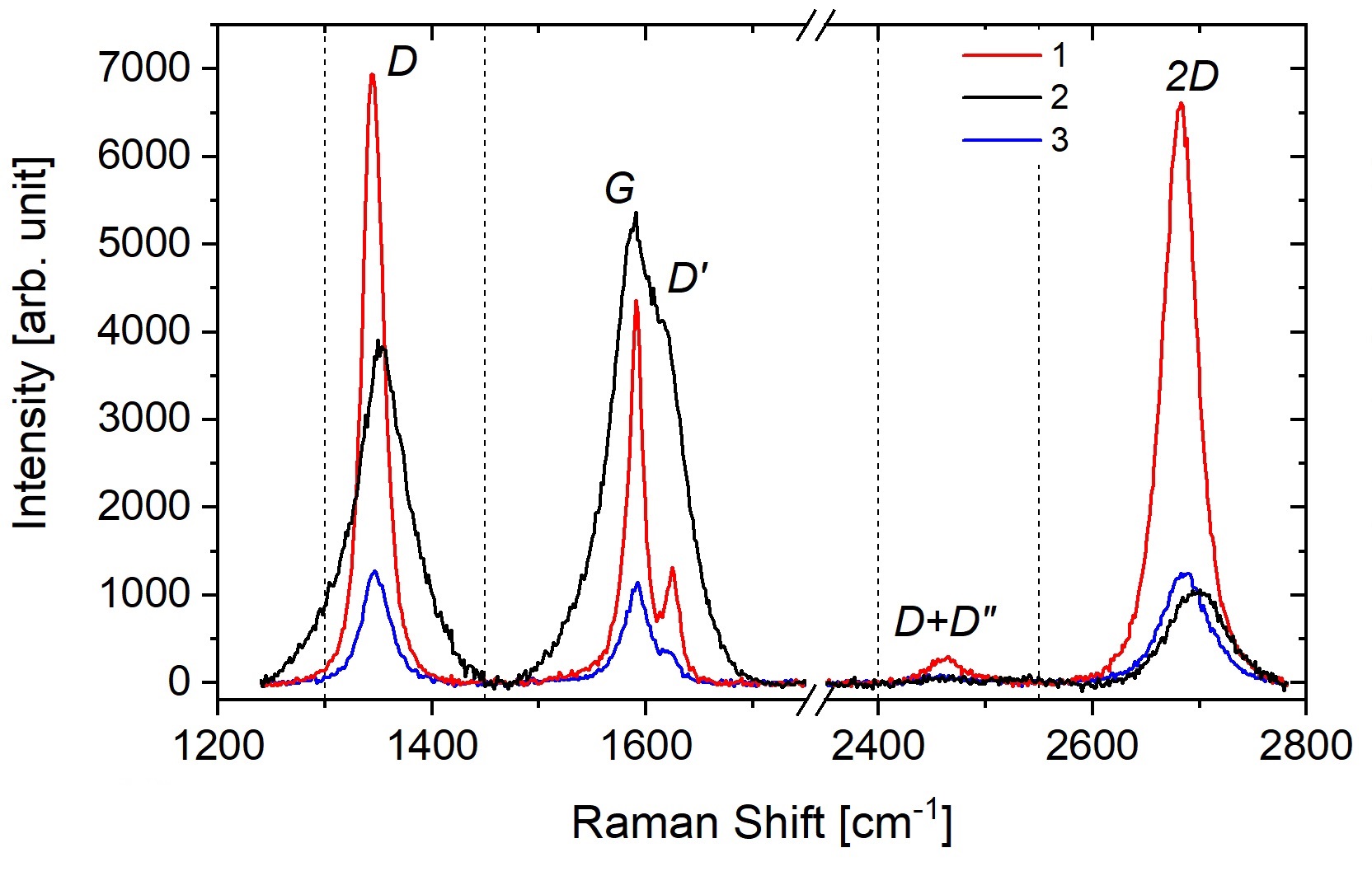}
\caption{Room temperature Raman spectroscopy measurements for graphene transferred from the Ni$_9$Mo$_1$(111) film taken at three different regions, showing the graphene's characteristic peaks. The dashed vertical lines indicate regions for the different peaks.}
\label{fig:RamanNiMoCu}
\end{figure*}

Single-layer graphene is known to have three characteristic features; \textit{I$_{2D}$}/\textit{I$_G$} ratio $>2$ and a 2\textit{D} peak fitted with a single Lorentzian with a full-width at half-maximum (FWHM) less than 40 cm$^{-1}$.\cite{Cabrero-Vilatela2016a} Furthermore, \textit{I}\textsubscript{\textit{D}}/\textit{I}\textsubscript{G} is usually used as an indicator of defects present in graphene, which increases with the increase of disorder in the graphene structure.\cite{Shlimak2015a,Cabrero-Vilatela2016a} Therefore, good-quality graphene possesses a small \textit{I$_D$}/\textit{I$_G$} ratio. Also, the \textit{D} peak was reported to change in shape, position and width by increasing the number of graphene layers, while the \textit{G} peak is downshifted with increasing the number of graphene layers, but upshifted with increasing the doping level.\cite{Malard2009} Moreover, the width of the 2\textit{D} peak increases with the number of graphene layers.\cite{Ferrari2013,Ferrari2007,Reichardt2017,Ferrari2006} Therefore, we assess the quality of our transferred graphene based on these features. Figure \ref{fig:RamanNiMoCu} shows the RT Raman spectra for graphene transferred from the Ni$_9$Mo$_1$ film. The average peaks positions and the average 2\textit{D} FWHM of all the samples are listed in  Table \ref{RamanSummary}. 
 
The Raman spectra of the transferred graphene from Ni$_9$Mo$_1$ (Figure \ref{fig:RamanNiMoCu} (a)) show a high \textit{I\textsubscript{D'}}. Therefore, an argument similar to that used for the graphene transferred from Ni(111) can be applied here to explain the spectra. However, the wide spatial variation across the surface of the sample, which could be attributed to formation of occasional strong Mo$-$C bonding between the graphene and the Ni$_9$Mo$_1$ film, makes it difficult to assess the quality of the graphene grown on Ni$_9$Mo$_1$ based on the 2\textit{D} FWHM.

\begin{table*}[t!]
\caption{The results of the Raman measurements summarising the average peaks positions and the 2\textit{D} average FWHM for graphene transferred from rotated Ni, epitaxial Ni and Ni$_9$Mo$_1$ films.}
\centering
\begin{tabular}{c P{1.5cm} P{1.5cm} P{1.5cm} P{1.5cm} P{1.5cm} P{2.0cm} }
	\hline 
Sample & \textit{D} [cm$^{-1}$] & \textit{G} [cm$^{-1}$] & \textit{D'} [cm$^{-1}$] & \textit{D}+\textit{D"} [cm$^{-1}$] & 2\textit{D} [cm$^{-1}$] & 2\textit{D} FWHM [cm$^{-1}$] \\  \hline
Rotated Ni(111) & 1346.3 & 1590.7 & 1624.8 & 2464.2 & 2681.9 & 46.2 \\
Epitaxial Ni(111) & 1344.6 & 1590.7 & 1623.1 & 2462.6 & 2682.6 & 40.8 \\
Ni$_9$Mo$_1$(111) & 1346.3 & 1590.7 & 1619.9 & 2464.0 & 2682.6 & 39.6 \\
\hline
\end{tabular}
\label{RamanSummary}
\end{table*} 

\section{X-ray Magnetic Circular Dichroism (XMCD) Formulae}
The X-ray absorption amplitude ($\mu_{XAS}$) and the X-ray magnetic circular dichroism ($\mu_{XMCD}$) are expressed as:
\begin{equation}    \label{XAS}
    \mu_{\text{XAS}}=\dfrac{1}{2}(\mu_{+}+\mu_{-}),
\end{equation}
and
\begin{equation}    \label{XMCD}
    \mu_{\text{XMCD}}=\mu_{+}-\mu_{-},
\end{equation}
were $\mu_+$ and $\mu_{-}$ are the absorption coefficients for the right and left circularly polarized light, respectively, normalized to a common value.\cite{OBrien1994a}

According to the sum rules, XAS and XMCD spectra can be used to determine the 3\textit{d} orbital angular momentum $<L_z>$ and the spin angular momentum $<S_z>$ for the \textit{L}\textsubscript{2,3}-edge using the following expressions:\cite{Legros1998b,Chen1995b}
\begin{equation}    \label{m_o}
    <L_z>=-2n_h\cdot\dfrac{\int_{L_3+L_2}(\mu_+-\mu_-)dE}{\int_{L_3+L_2}(\mu_++\mu_-+\mu_{XAS})dE},
\end{equation}
\begin{equation}    \label{m_s}
\begin{split}
<S_z>=-\dfrac{3}{2}n_h\cdot\dfrac{\int_{L_3}(\mu_+-\mu_-)dE-2\int_{L_2}(\mu_+-\mu_-)dE} {\int_{L_3+L_2}(\mu_++\mu_-+\mu_{XAS})dE} \\ \cdot\Big(1+\dfrac{7<T_z>}{2<S_z>}\Big).
\end{split}
\end{equation}
Here, $n_h=(10-n_d)$ is the number of holes in the 3\textit{d} states, whereas $n_d$ is the electron occupation number of the 3\textit{d} states. $L_3+L_2$ represent the integration range over the \textit{L}$_3$ and \textit{L}$_2$ edges and $<T_z>$, which is the expectation value of the magnetic dipole operator, is known to be very small in TMs and hence it was ignored.\cite{Idzerda1995} 
In (\ref{XMCD}), $\mu$\textsubscript{XMCD} is corrected for the light degree of polarization, $P$, and the light's incident angle, $\theta$. Therefore, it was multiplied by a factor $\dfrac{1}{P\cos\theta}$,\cite{Amemiya2001} where $\theta$ is measured with respect to the sample's surface, while $\mu$\textsubscript{XAS} remains unchanged.\cite{Sun2014a} 

The orbital magnetic moment, $m_o$, is equal to $\mu_{\textrm{B}}<L_z>$, whereas the spin magnetic moment, $m_s$, is  given by $m_s=2\mu_{\textrm{B}}<S_z>$.\cite{Chen1995b} For the total magnetic moment, $m_{total}=m_o+m_s$.\cite{Wende2007b,Vogel1997,Chen1995b}

In contrast, since the XMCD at the \textit{K}-edge measures the transitions from a non-spin-split \textit{s} orbital to \textit{p} orbital, only $m_o$ can be obtained:\cite{Vita2014,Huang2002}
\begin{equation}
m_o=-\dfrac{1}{3}\dfrac{n_h}{P\cos\theta}\dfrac{\int_K\mu_{XMCD}}{\int_K\mu_{XAS}}, \label{m_o K-edge}
\end{equation}

For the \textit{K}-edge, $n_h$ becomes equal to $6-n_p$, where $n_p$ is the electron occupation number in the 2\textit{p} bands.\cite{Huang2002} Since $m_s=2m_o/(g-2)$, where $g$ is the gyromagnetic factor,\cite{Pelzl2003} $m_{total}$ can be estimated for the \textit{K}-edge if $g$ is known.

\section{Polarized Neutron Reflectometry Analysis} \label{SI:PNR}

The primary analysis of the PNR data was performed using the \textit{Refl1D} software package from the National Center for Neutron Research (NCNR) at the National Institute for Standards and Technology (NIST).\cite{Refl1d_URL_ref} \textit{Refl1D} uses the Bayesian analysis fitting package Bumps,\cite{Bumps_URL_ref} for both obtaining the best fit and performing an uncertainty analysis on the parameters used to model the data. \textit{Refl1D} is based on the Abeles optical matrix method\cite{Abeles_optical_matrix_method} along with interfaces approximated using the method of N\'{e}vot and Croce.\cite{Nevot_RevuedePhysAppl_1980} 
\textit{Refl1D} allows multiple data sets to be fitted simultaneously which is advantageous in this case, due to the difficulty of detecting such a thin layer of graphene on top of a thick layer of Ni metal. Hence the data taken at the two temperatures (10 K and 300 K) were fitted simultaneously for each sample, to reduce the uncertainly in the fitted parameters, such as the magnetic contrast variation. This is akin to the isotopic contrast variation method used as standard in non-polarized neutron reflectivity soft matter experiments.\cite{Koutsioubas:kc5089}

\subsection{Modeling methodology}

The analysis was started on the rotated-domain graphene/Ni sample using the simplest model possible as a starting point. Once convergence was reached as defined in the \textit{Refl1D} documentation,\cite{Refl1d_convergence} an assessment was made taking into account the improvement of the $\chi^{2}$ and the uncertainty in the parameters as to whether to increase the complexity of the model by adding more parameters. Prior knowledge of the sample growth conditions and sample characterisation was also employed to aid in this process.  
We echo what is said in the \textit{Refl1D} documentation\cite{Refl1d_convergence} that, since this is parametric modeling technique, we cannot say that any one model is truly correct, only that while it fits the observed data well, there may well be other possible solutions. We couple this knowledge with the following secondary information:

\begin{itemize}
  \item A well defined question for what the modeling is intended to answer. This provides a measure of how well the model is working, but most importantly, a point at which to stop fitting the data. It can also be instrumental in determining if the data quality is insufficient to answer the posed question.
  \item A knowledge of the sample growth conditions to provide, initial priors (maximum and minimum bounds on those initial parameters) and a nominal starting structure. In particular, the substrate material parameters should be well known.
  \item Magnetic contrast variation, which is akin to solving simultaneous equations.
  \item Secondary characterization techniques to provide a cross reference for a particular parameter and lock them down, e.g. XMCD for the total moment. This can also be crucial in dealing with issues ultimately arising from the inverse phase problem, such as multiple solutions with similar $\chi^{2}$, i.e. multi-modal fit solutions. 
\end{itemize}

Such information provides confidence in the validity of the model and avoid over parameterization. 

In the present case, the question is: can the model confirm that there is a graphene layer present  and primarily is it magnetic and can the magnitude of the magnetic moment be measured with any certainty?  

\begin{figure*}
	\includegraphics[width=1.0\textwidth]{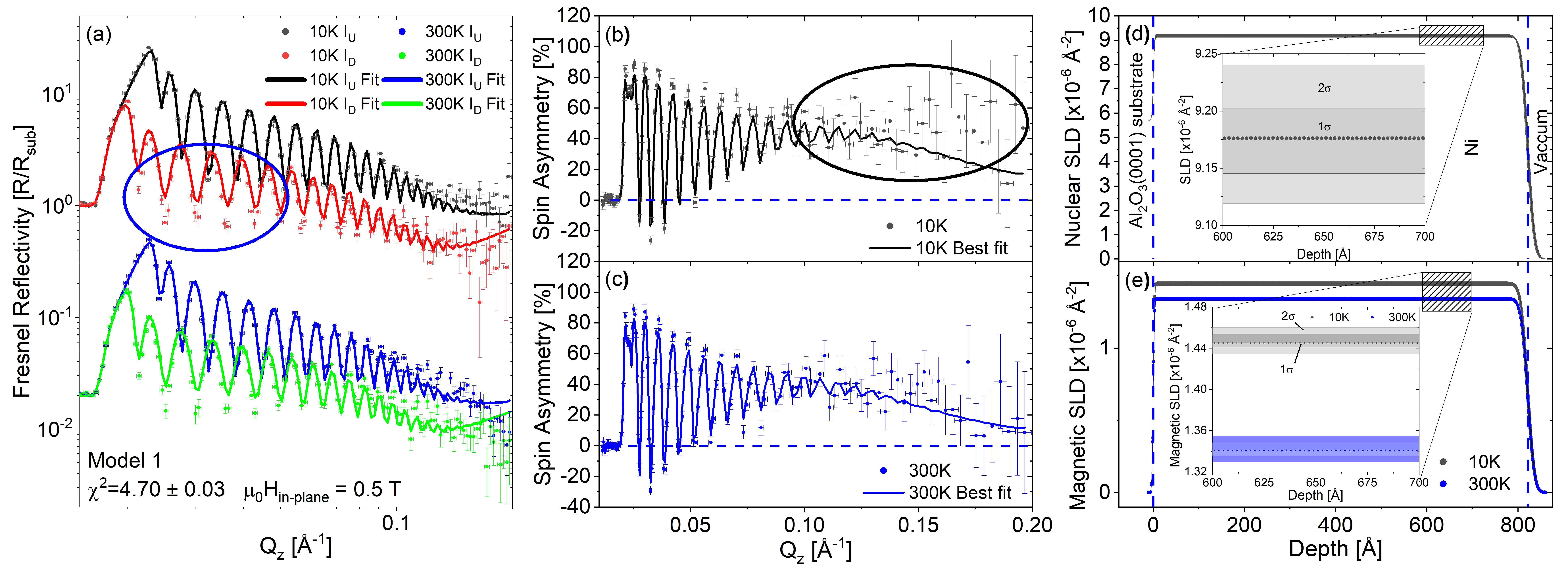}
	\caption{\label{PNR1}
		PNR Model 1: Single Ni layer with commensurate magnetism. (a) Fresnel reflectivity for 10 K and 300 K,  (b) and (c) show the spin asymmetries, (d) is the nuclear scattering length density (nSLD) profile and (e) the magnetic scattering length density (mSLD) profile. The grey banded regions around the SLD lines are the 95\% Bayesian confidence intervals.
	}
\end{figure*}

Model 1 consists of a single layer of Ni on an Al$_{2}$O$_{3}$ substrate with the Ni magnetism commensurate with the boundaries of the Ni Layer. There are no magnetic dead-layers included and no graphene layer. The aim is to provide a null result to see if the graphene is required to fit the dataset. The results are shown in Figure \ref{PNR1}. Panel (a) displays the Fresnel reflectively (\textit{R\textsubscript{Fresnel}}) given by: 

\begin{equation}
  \textrm{\textit{R}}_{\textrm{\textit{Fresnel}}}=\frac{\textrm{\textit{R}}}{\textrm{\textit{R}}_{\textrm{\textit{Substrate}}}},
\end{equation}

which is the total reflectivity (\textit{R}) divided by the calculated substrate reflectivity (\textit{R$_{Substrate}$}). This visually aids finding discrepancies between the fit and the data. In this case it is obvious that the fit is not matching the depth of the fringes at low Q as ringed in blue. 

%
%
Panels (b) and (c) show the spin asymmetries ($SA$). It is clear the fit does not fit well magnetically at high \textit{Q} for both 10 K and 300 K. This is in the same region in the \textit{R\textsubscript{Fresnel}} that curves up at high $Q$. This is indicative of a thin and magnetic layer being missing from the model, such as a thin graphene layer. The oscillations also don't match at low \textit{Q}. Panel (d) shows the nuclear scattering length density (nSLD) profile. The inset showing the 68\% and 95\% Bayesian confidence intervals from the Bumps fitting package. This gives some indication of how certain the model is with regards to the fit of the data, but we stress that this just indicates how well the current model fits the data, it does not indicate if the model is the correct one or not.

\begin{figure*}
	\includegraphics[width=1.0\textwidth]{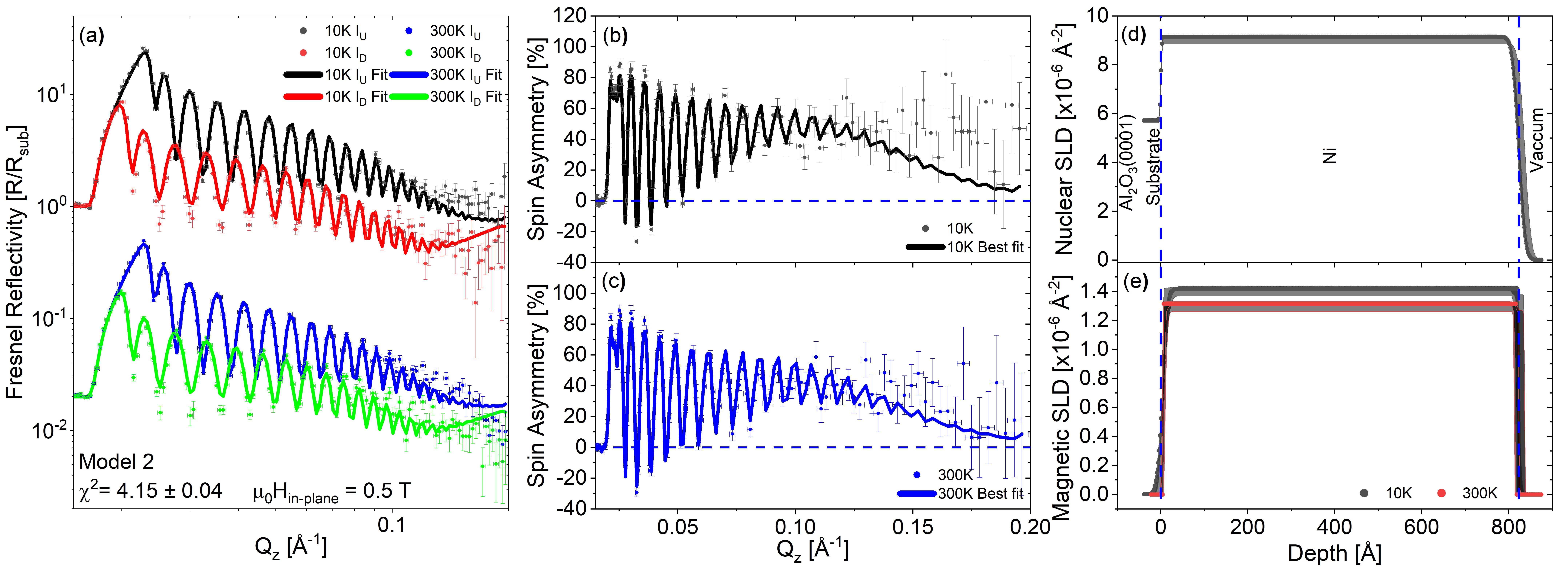}
	\caption{\label{PNR2}
		PNR Model 2: Ni layer with dead-layers top and bottom. (a) Fresnel reflectivity for 10 K and 300 K,  (b) and (c) show the spin asymmetries, (d) is the nuclear scattering length density (nSLD) profile and (e) the magnetic scattering length density (mSLD) profile. The grey banded regions around the SLD lines are the 95\% Bayesian confidence intervals.
	} 
\end{figure*}

It follows from model 1 that a thin layer or layers is required to fit the high $Q$ data. This is not necessarily a graphene layer, it could be a thin magnetic dead layer. Therefore we constructed model 2 shown in Figure \ref{PNR2}. This model consists of a single layer of Ni on an Al$_{2}$O$_{3}$ substrate where the Ni magnetism is now incommensurate with the boundaries of the Ni layer, i.e. we allow Ni dead-layers with a moment that can vary from zero to the full Ni moment and have thicknesses and roughnesses independent of the nSLD profile. Again this is intended as a null case. This model slightly improves the $\chi^{2}$ however still fails to fit the low and high $Q$ regions that model 1 did not reproduce.
\begin{figure*}
	\includegraphics[width=1.0\textwidth]{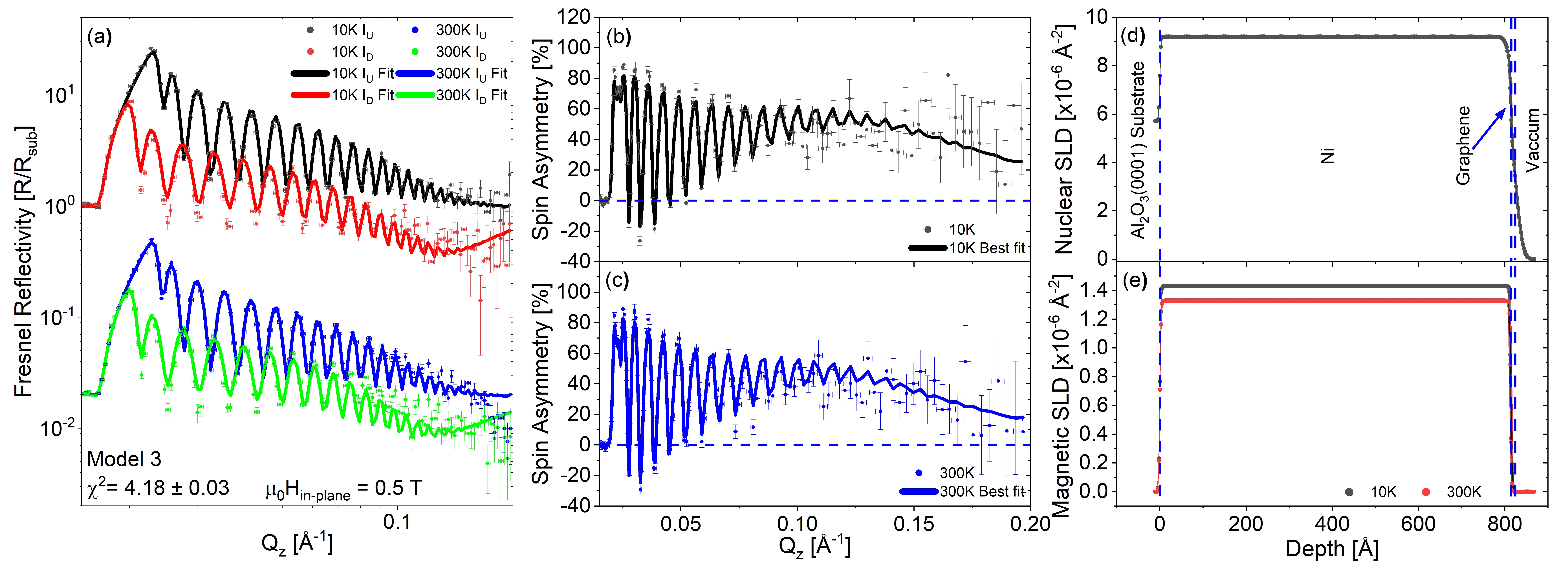}
	\caption{\label{PNR3}
    		PNR Model 3: Ni layer with commensurate magnetism and a nonmagnetic graphene layer on the surface. (a) Fresnel reflectivity for 10 K and 300 K,  (b) and (c) show the spin asymmetries, (d) is the nuclear scattering length density (nSLD) profile and (e) the magnetic scattering length density (mSLD) profile. The grey banded regions around the SLD lines are the 95\% Bayesian confidence intervals.
    	} 
\end{figure*}

Model 3, shown in Figure \ref{PNR3}, consists of a single layer of Ni on an Al$_{2}$O$_{3}$ substrate with the Ni magnetism commensurate with the boundaries of the Ni Layer and a non-magnetic graphene layer on the top surface. This produces a fit with a similar $\chi^{2}$ to model 2 that still fails to sort the issues at high and low $Q$ from the previous models. 

\begin{figure*}
	\includegraphics[width=1.0\textwidth]{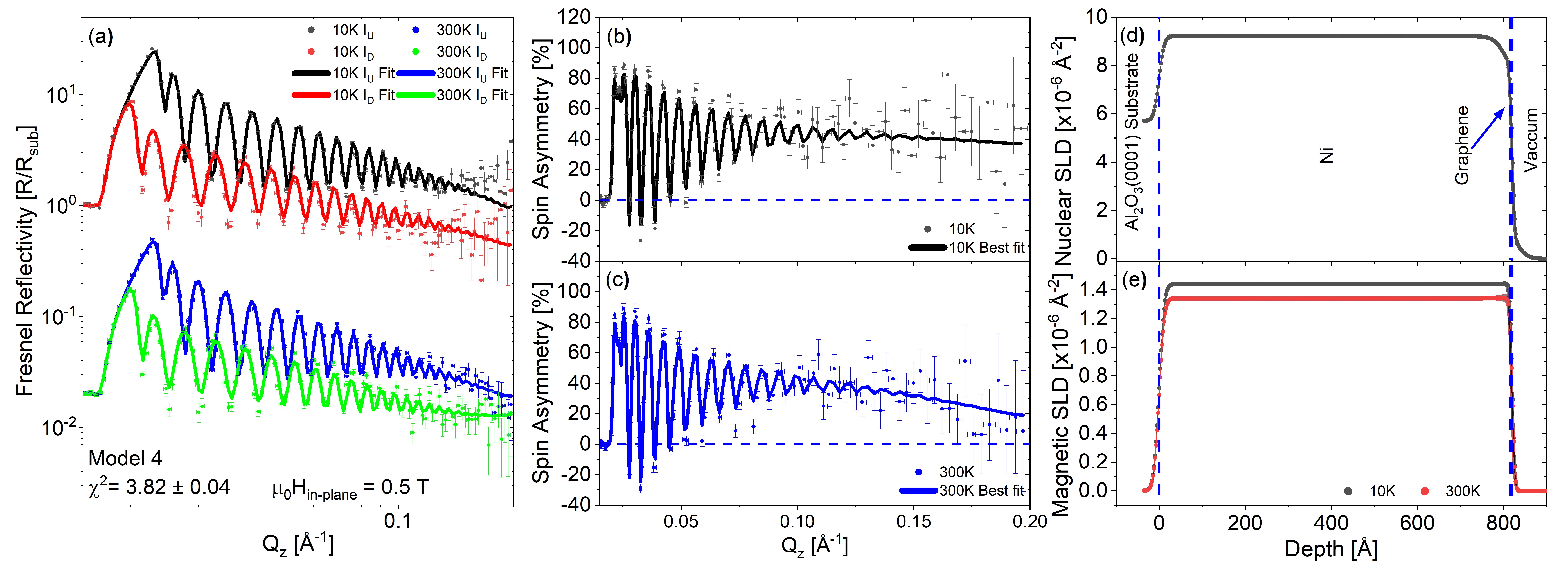}
	\caption{\label{PNR4}
		PNR Model 4: Ni layer with commensurate magnetism and a magnetic graphene layer on the surface. (a) Fresnel reflectivity for 10 K and 300 K,  (b) and (c) show the spin asymmetries, (d) is the nuclear scattering length density (nSLD) profile and (e) the magnetic scattering length density (mSLD) profile. The grey banded regions around the SLD lines are the 95\% Bayesian confidence intervals.
	}
\end{figure*}

The next step is to allow the graphene layer to be magnetic as shown in Figure \ref{PNR4} for model 4, while keep the rest of the model the same as model 3. Although this model produces a small step change in the $\chi^2$ where the high $Q$ data is significantly improved, it does not capture the low $Q$ fringes. However, it provides some evidence that we need a graphene layer that is magnetic to fit the data. At this point it seems likely that some combination of Ni dead-layers and magnetic graphene maybe at play, however, certainty on the values eludes us since the fit cannot reproduce the low $Q$ data. 

\begin{figure*}[t!]
	\includegraphics[width=1.0\textwidth]{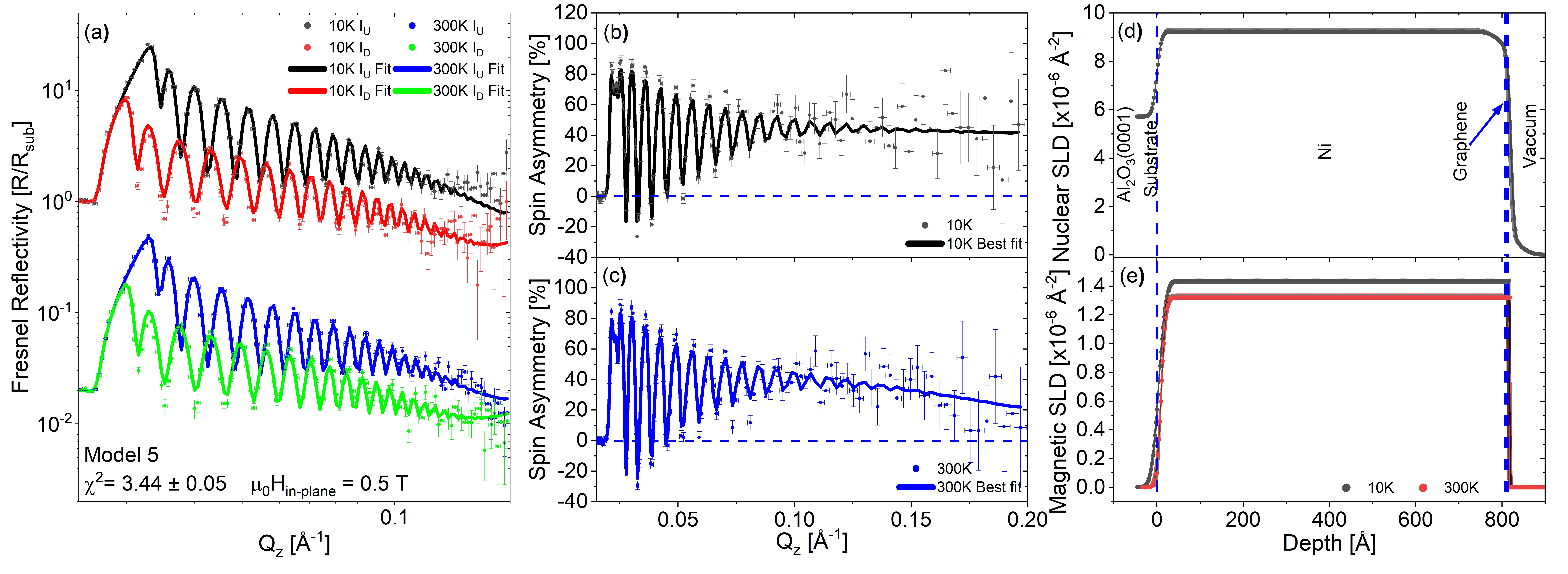}
	\caption{\label{PNR5}
		PNR Model 5: Ni layer with dead-layers top and bottom a nonmagnetic graphene layer on the surface. (a) Fresnel reflectivity for 10 K and 300 K, (b) and (c) show the spin asymmetries, (d) is the nuclear scattering length density (nSLD) profile and (e) the magnetic scattering length density (mSLD) profile. The grey banded regions around the SLD lines are the 95\% Bayesian confidence intervals.
	} 
\end{figure*}

Model 5, shown in Figure \ref{PNR5}, is another null model, where the statistics dominate, consisting of a single layer of Ni on an Al$_{2}$O$_{3}$ substrate with the Ni magnetism having dead layers top and bottom with a non-magnetic graphene layer at the surface. This produces a significant improvement in $\chi^{2}$ as you would expect from the earlier model, hinting that some combination is required. 

\begin{figure*}[t!]
	\includegraphics[width=1.0\textwidth]{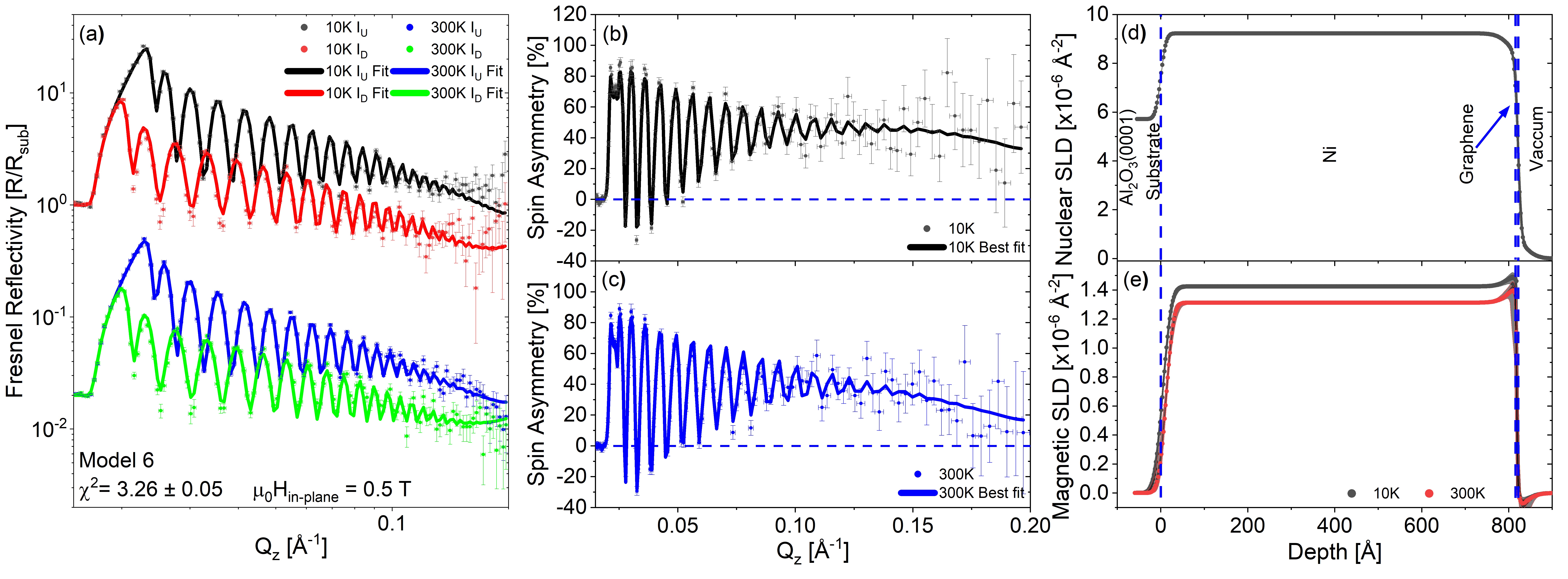}
	\caption{\label{PNR6}
    		PNR Model 6: Ni layer with dead-layers top and bottom and a magnetic graphene layer on the surface. (a) Fresnel reflectivity for 10 K and 300 K,  (b) and (c) show the spin asymmetries, (d) is the nuclear scattering length density (nSLD) profile and (e) the magnetic scattering length density (mSLD) profile. The grey banded regions around the SLD lines are the 95\% Bayesian confidence intervals.
    	} 
\end{figure*}

Model 6 takes model 5 and allows the graphene to be magnetic. This results in a magnetic discrepancy at the top interface due to a small region of magnetically dead Ni leading to magnetic graphene as shown in the mSLD in Figure \ref{PNR6} (e). This manifests as a spike in the magnetization of the graphene layer and is not physical.

\begin{figure*}[t!]
	\includegraphics[width=1.0\textwidth]{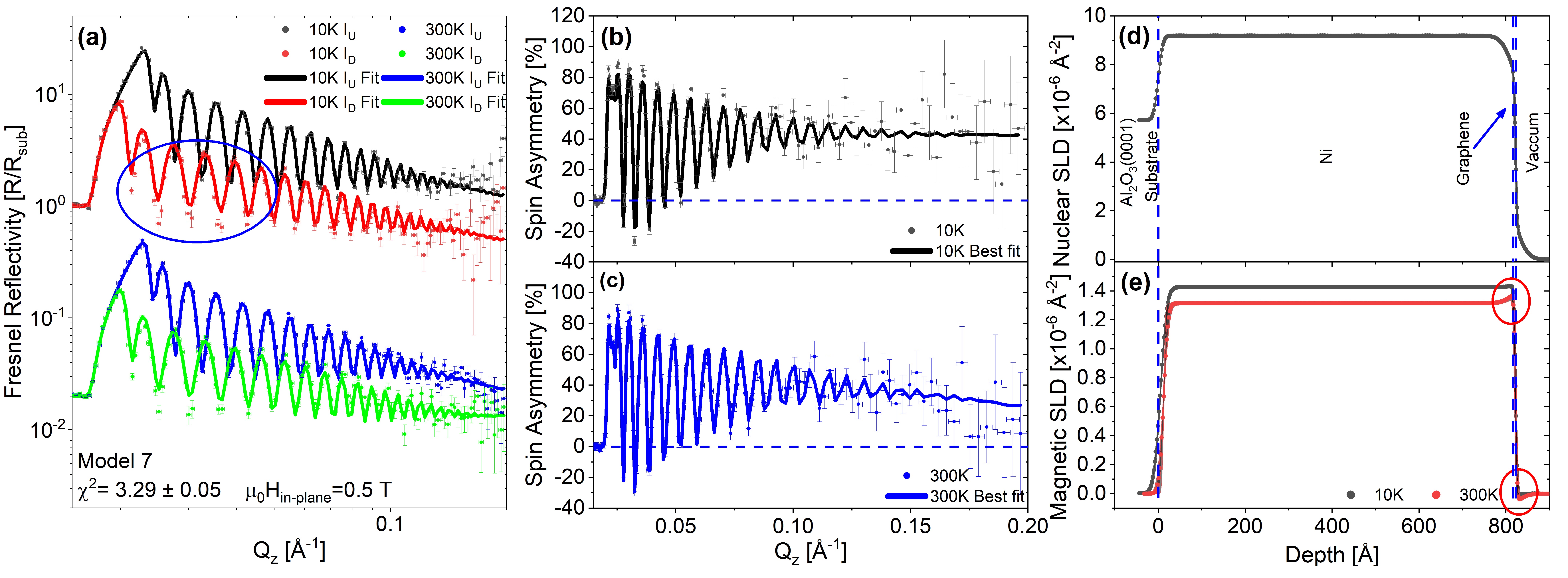}
	\caption{\label{PNR7}
		PNR Model 7: Ni layer with a dead layer at the substrate interface and continuous magnetism into a magnetic graphene layer on the surface. (a) Fresnel reflectivity for 10 K and 300 K, (b) and (c) show the spin asymmetries, (d) is the nuclear scattering length density (nSLD) profile and (e) the magnetic scattering length density (mSLD) profile. The grey banded regions around the SLD lines are the 95\% Bayesian confidence intervals.
	}
\end{figure*}

Model 7, presented in Figure \ref{PNR7}, removes the top Ni dead-layer allowing continuous magnetism into the graphene. Although this model does not improve the $\chi^{2}$ over model 6, it slightly improves the top interface. However, there are still issues highlighted by red circles in panels (d) and (e) where this time the Ni roughness violates the N\'evot-Croce\cite{Nevot_RevuedePhysAppl_1980} approximation and exceeds the thickness of the graphene. This produces non-physical magnetic profiles, with a peak and dip at the interface regions. This is also present in Model 6. At this point it is obvious that the fit is limited by the low $Q$ features ringed in blue in Figure \ref{PNR7}. The fit cannot improve at high $Q$ unless the low $Q$ is improved due to the weighting of the statistics.

\begin{figure*}[t!]
	\includegraphics[width=1.0\textwidth]{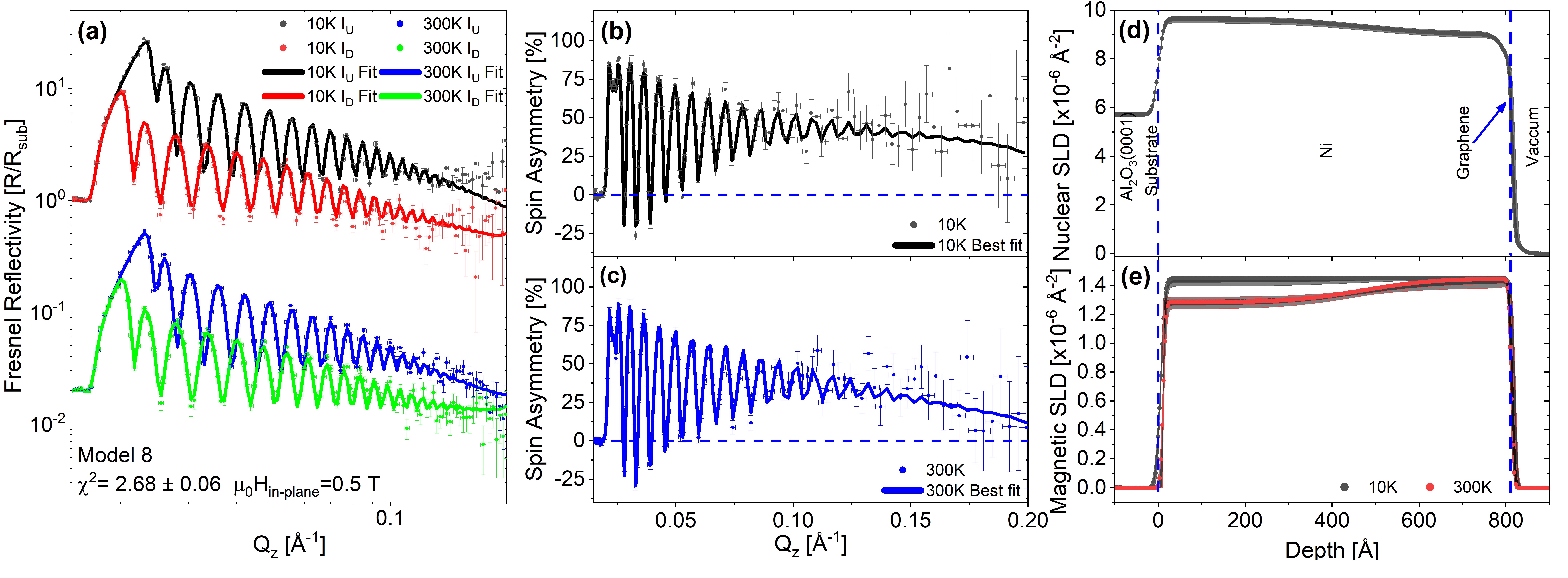}
	\caption{\label{PNR8}
		PNR Model 8: Ni layer with a dead layer at the substrate interface and continuous magnetism into a magnetic graphene layer on the surface. Graphene layer thickness is unconstrained. (a) Fresnel reflectivity for 10 K and 300 K, (b) and (c) show the spin asymmetries, (d) is the nuclear scattering length density (nSLD) profile and (e) the magnetic scattering length density (mSLD) profile. The grey banded regions around the SLD lines are the 95\% Bayesian confidence intervals.
	} 
\end{figure*}

Models $1- 7$ show that small changes to the graphene and magnetic dead-layers have no effect on the low $Q$ oscillations whose statistics are dominating the $\chi^{2}$ and so prevent a good fit of the high $Q$. It is noteworthy to mention that high $Q$ fringes contain information on the graphene layer, whereas that at low $Q$ range provide details about the Ni film. Model 8 (Figure \ref{PNR8}) takes model 7 and splits the Ni layer into two, allowing a large roughness between the two layers to grade the nSLD and mSLD, while the two are kept magnetically linked. Having two very similar nSLD layers allows some extra contrast and beat frequencies that can modify the amplitude of the Kiessig oscillations. This dramatically improves the $\chi^{2}$ and the uncertainties on the graphene layer as the fit is no longer dominated by the residuals at low $Q$. We have little justification for this, but as the Ni layer is relatively thick, if there is a clamping to the Al$_{2}$O$_{3}$ substrate then the approx 80 nm Ni(111) layer may relax its stress/strain across this thickness, producing an nSLD gradient across the layer.

\begin{figure*}[t!]
	\includegraphics[width=1.0\textwidth]{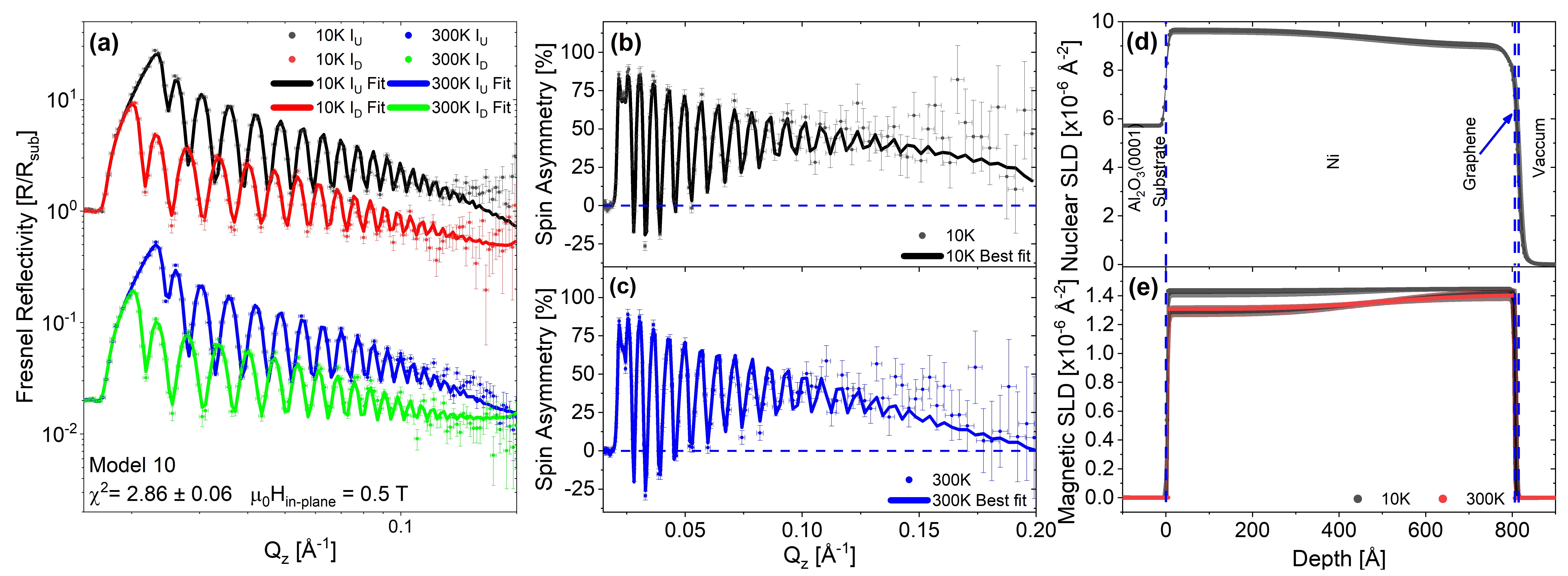}
	\caption{\label{PNR10}
		PNR Model 10: Ni layer with a dead layer at the substrate interface and continuous magnetism into a magnetic graphene layer on the surface. The Ni has been split into two continuous Ni layers to allow a gradient across the layer. The graphene layer is constrained so it cannot be thinner than 2 monolayers of graphene inline with the SEM and Raman results. It has also been set to be non-magnetic. (a) Fresnel reflectivity for 10 K and 300 K, (b) and (c) show the spin asymmetries, (d) is the nuclear scattering length density (nSLD) profile and (e) the magnetic scattering length density (mSLD) profile. The grey banded regions around the SLD lines are the 95\% Bayesian confidence intervals.
	} 
\end{figure*}

Model 8 is a workable fit demonstrating  magnetism in the graphene layer, however the graphene layer thickness reduces down to $\sim$ 2 \AA$\,$, which is right at the fitting limit set for one monolayer. The uncertainty in the fitted parameter indicates that we are not very sensitive to it. At this point we have not included any information from other characterization techniques. Both SEM and Raman spectroscopy measurements indicate that the Ni(111) with rotated-domain graphene should have approximately two to three monolayers of graphene on the surface. Therefore, if we use this information to set the lower bound on the prior for the graphene thickness, we get model 9 (shown in Figure 3 of the main text). This produces a fit within the error of the $\chi^{2}$ of model 8 but utilising all the information we have available. 

Four further checks were performed to study the sensitivity of the PNR modeling to the magnetic moment in the graphene layer. This is now a crucial check as all the fits discussed previously (i.e. before model 9) have been dominated by the residuals at low $Q$. These were not adequately modelled until the Ni layer was allowed to be graded. Therefore, model 10 (Figure \ref{PNR10}), is based on model 9, but the magnetic moment in graphene is fixed to zero. The fit produced a $\chi^2$ which is only slightly worse than model 9, see Table \ref{table3}. This indicates we still have little sensitivity to any magnetism in the graphene.

\begin{figure*}[t!]
	\includegraphics[width=17.2cm]{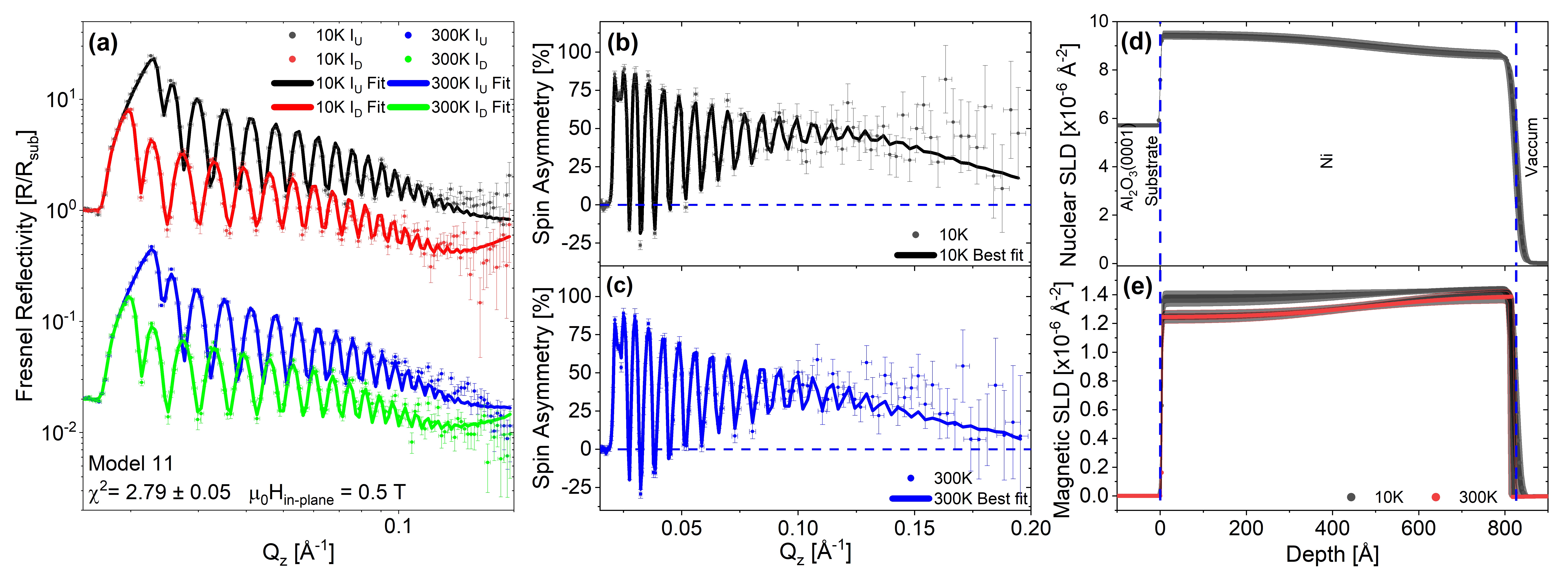}
	\caption{\label{PNR11}
		PNR Model 11: Ni layer has dead layers at the substrate interface and top Ni/Graphene interface. The Ni has been split into two continuous Ni layers to allow a gradient across the layer. The graphene layer has been removed. (a) Fresnel reflectivity for 10 K and 300 K,  (b) and (c) show the spin asymmetries, (d) is the nuclear scattering length density (nSLD) profile and (e) the magnetic scattering length density (mSLD) profile. The grey banded regions around the SLD lines are the 95\% Bayesian confidence intervals.
	} 
\end{figure*}

The next test is where the graphene layer is removed entirely, the results of which are shown in Figure \ref{PNR11}. Again, the magnetism in the Ni has dead layers top and bottom. Surprisingly, this almost matches Model 9's $\chi^{2}$, being fractionally worse but with overlapping error bars. 

\begin{figure*}[t!]
	\includegraphics[width=17.2cm]{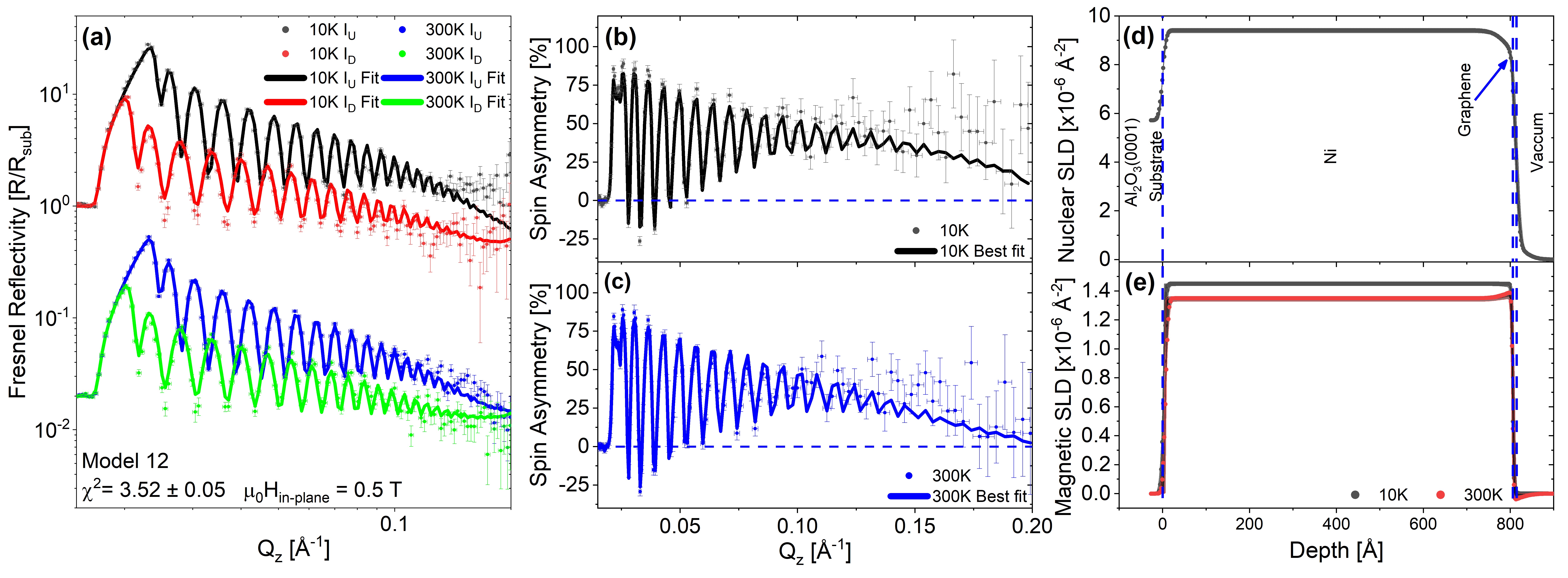}
	\caption{\label{PNR12}
		PNR Model 12: a less realistic version of Model 10 for completeness. The Ni layer has a dead layer at the substrate interface only. The Ni has been split into two continuous Ni layers to allow a gradient across the layer. The graphene layer is constrained so it cannot be thinner than 2 monolayers of graphene inline with the SEM and Raman results. The graphene is set to be non-magnetic. (a) Fresnel reflectivity for 10 K and 300 K,  (b) and (c) show the spin asymmetries, (d) is the nuclear scattering length density (nSLD) profile and (e) the magnetic scattering length density (mSLD) profile. The grey banded regions around the SLD lines are the 95\% Bayesian confidence intervals.
	} 
\end{figure*}

\begin{figure*}[t!]
	\includegraphics[width=17.2cm]{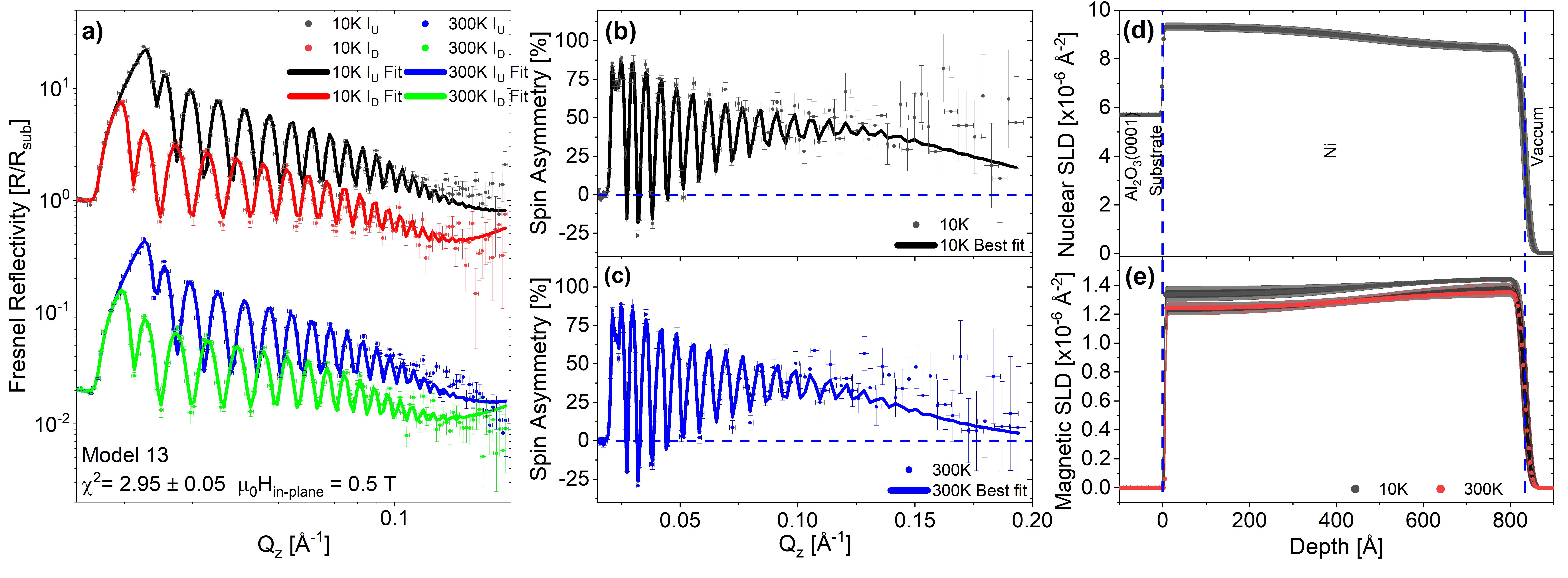}
	\caption{\label{PNR13}
		PNR Model 13: a less realistic version of Model 11. The Ni layer has a dead layer at the substrate interface only. The Ni has been split into two continuous Ni layers to allow a gradient across the layer, and the graphene layer has been removed. (a) Fresnel reflectivity for 10 K and 300 K,  (b) and (c) show the spin asymmetries, (d) is the nuclear scattering length density (nSLD) profile and (e) the magnetic scattering length density (mSLD) profile. The grey banded regions around the SLD lines are the 95\% Bayesian confidence intervals.
	} 
\end{figure*}

Models 12 and 13 shown in Figures \ref{PNR12} and \ref{PNR13} are copies of models 10 and 11, respectively, but with dead layers at the bottom interfaces only. The magnetism at the top is conformal with the nuclear structure as in model 9. Both of which produce worse fits than model 9, and lead to the idea that either a graphene layer or a top magnetic dead-layer in the Ni is needed to improve the fit. 

\begin{table*}[t]
\centering
\caption{Comparison of model parameters, $\chi^{2}$ values and Bayesian evidence terms.}
 \begin{tabular}{P{2.2cm}P{3.0cm}P{2.0cm}P{2.2cm}P{2.4cm}P{2.4cm}}
  \hline
  Model Number & Number of Parameters & $\chi^{2}$ & $\chi^{2}$ error & Logz (Evidence) & $Logz_err$\\
  \hline
    $9$ & $28$ & $2.73$ & $\pm\; 0.06$ & $-868.0$ & $\pm\; 0.3$\\
    10 & 30 & 2.86 & $\pm$ 0.06 & -915.0 & $\pm$ 0.7\\
    11 & 27 & 2.79 & $\pm$ 0.05 & -1097.6 & $\pm$ 0.5\\
    12 & 26 & 3.52 & $\pm$ 0.05 & -888.7 & $\pm$ 0.4\\
    13 & 23 & 2.95 & $\pm$ 0.05 & -927.1 & $\pm$ 0.4\\
  \hline
  \multicolumn{6}{ c }{Table of Evidence} \\
 \end{tabular}
\label{table3}
\end{table*}
We are faced with the conundrum of now having 5 models all with very similar $\chi^{2}$ values. Taking into account the extra evidence provided by the SEM, Raman and XMCD allows us to qualitatively select the model with two monolayers of graphene that are magnetic. Ideally, we would like a quantitative measure applicable to the PNR fits to also show this is the most likely case.

It is possible to obtain a more quantitative measure by refitting the models using a nested sampler.\cite{doi:10.1063/1.1835238,buchner2021nested} Here we used the UltraNest package.\cite{2021JOSS....6.3001B}

The details of how nested sampling works are beyond the scope of this work and we direct the reader to the work of Skilling $et$ $al$\cite{doi:10.1063/1.1835238} who developed nested sampling as a method of Bayesian evidence determination. The work of McCluskey $et$ $al$\cite{McCluskey_2020} deploys nested sampling for analysing neutron reflectivity data for a soft matter systems, where they obtained the Bayesian evidence term and used it to compare different complexities of model and which most meaningfully fit the data.

This approach avoids the risk of over-fitting as the Bayesian evidence is derived from an integral in parameter space and therefore scales with the number of parameters. It is akin to having a built-in Occam's razor which means that any additional free parameters to the model must significantly improve the likelihood of the determined value. However, it is important also to note that the accuracy of the determined evidence depends on the prior probabilities chosen for each of the free parameters and therefore care should be taken to ensure that these are meaningful.\cite{McCluskey_2020} All shared parameters between the models have strictly the same priors (uniform ranges and distribution). It is the setting of new additional parameter priors, in our case the graphene thickness and magnetism, as maximum and minimum ranges (assuming a uniform probability distribution as in our case) via our secondary evidence (SEM, Raman and XMCD) that allows us to get some measure of how including this information in the fit gives some validity to the models even when they have very similar $\chi^2$ values. Hence, the Bayesian evidence allows for the comparison of two models, with different numbers of parameters, given that the data they are applied to is the same.

The evidence term is outputted as a negative log-likelihood.\cite{doi:10.1063/1.1835238} In this case, the larger the number is, the more probable the model is. For example, -2 is more probable than -5. There are ways of further utilising the evidence term but we refer the reader to the literature with regards to these.\cite{doi:10.1080/01621459.1995.10476572,10.3389/fpsyg.2019.02767}

\begin{figure*}[t!]
	\includegraphics[width=17.2cm]{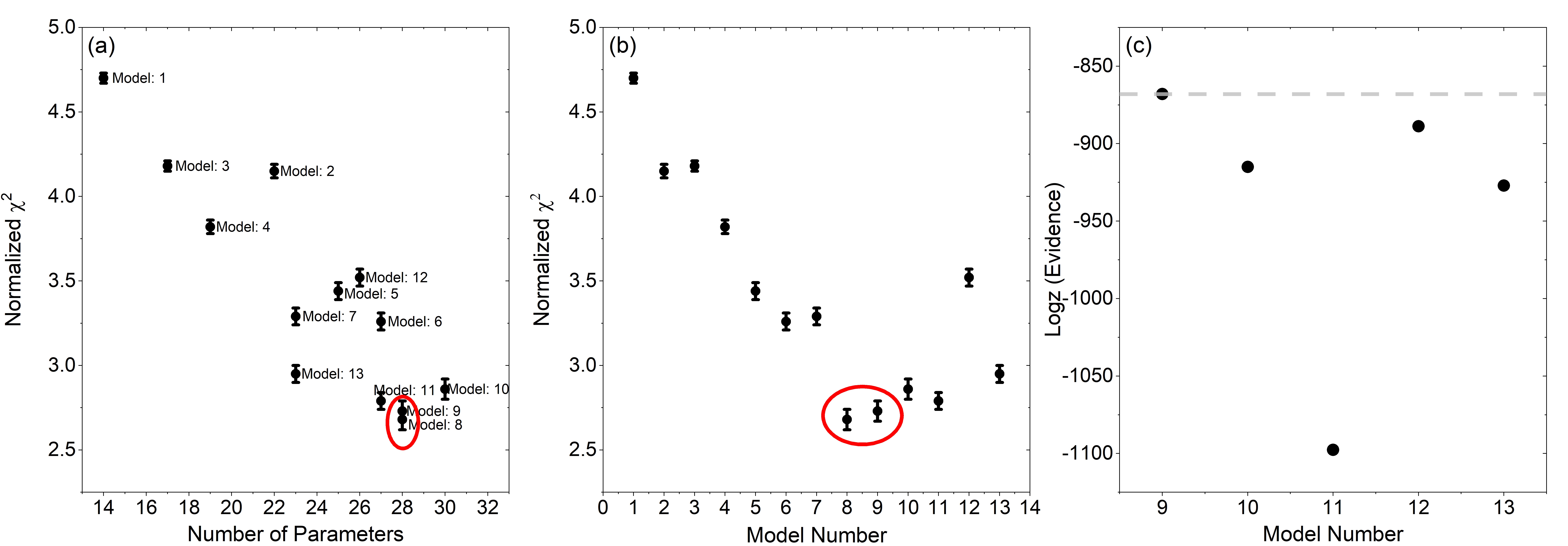}
	\caption{\label{PNR14}
		Panels (a) and (b) show the trends in $\chi^{2}$ vs the number of parameters and model number, respectively. Models 8 and 9 have the lowest (best) $\chi^{2}$ values. However, Model 8 is dismissed as the thickness of the graphene was allowed to vary freely and shrunk to 2\AA, which is not physical. The Raman and SEM results indicate that two monolayers of graphene are present. (c) Bayesian evidence term as computed by the UltraNest nested sampler package.\cite{2021JOSS....6.3001B} The highest values has the greatest evidence, the dashed grey line indicates model 9's cut-off vs the 4 main closest physical models with similar $\chi^{2}$. These models (10-13) all dismiss the secondary information from the SEM, Raman and XMCD, that show there are two monolayers of graphene layer present and that it is magnetic in nature. The error bars in panel (c) are smaller than the data points. 
	} 
\end{figure*}

\begin{figure*}[t!]
	\includegraphics[width=17.2cm]{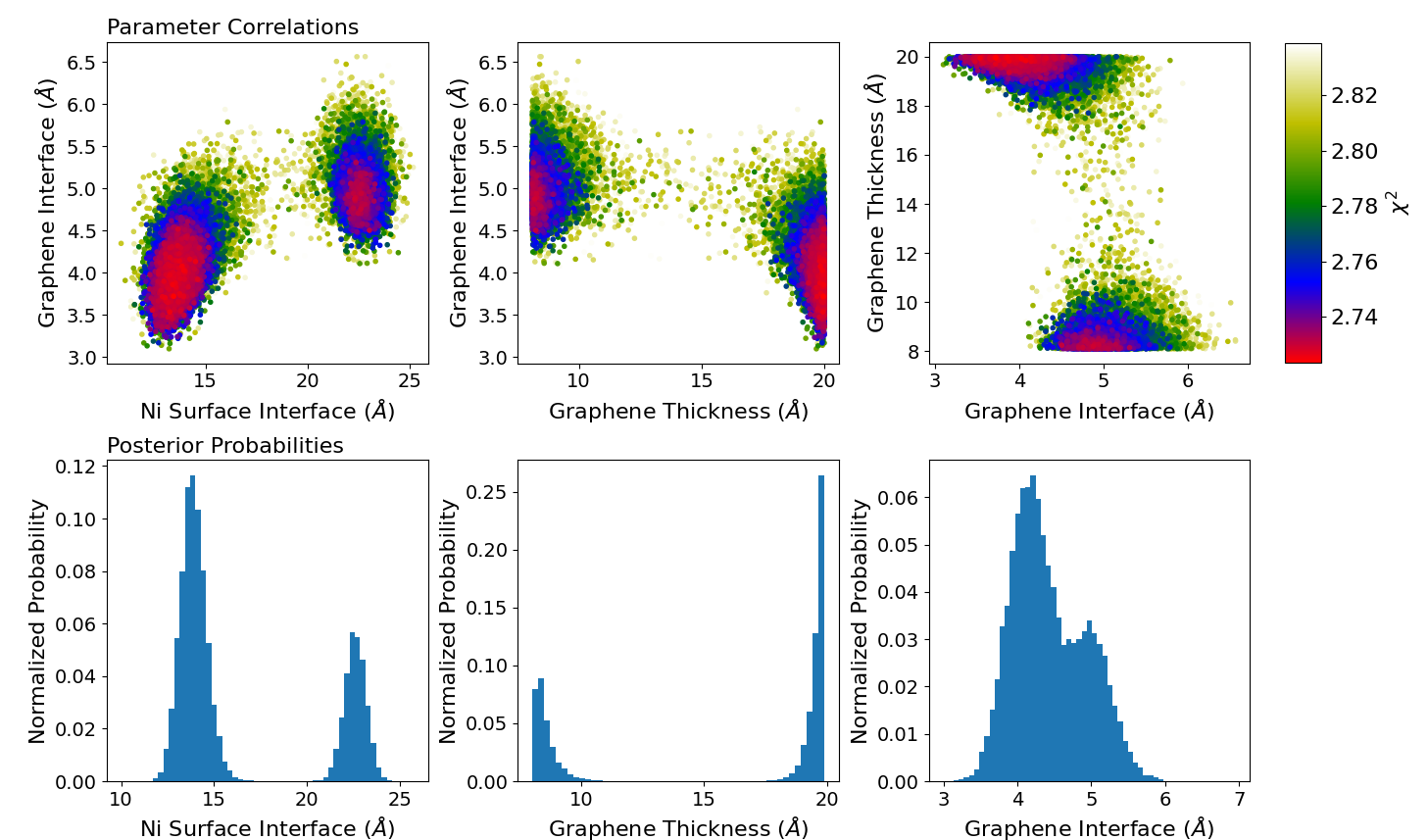}
	\caption{\label{PNR15}
		Correlation plots and posterior probability distributions for the parameters exhibiting bi-modal behaviour from Model 9. Secondary information (obtained from the SEM and Raman spectroscopy measurements) was used to select one mode as preferable in each case.
	} 
\end{figure*}

Models 9-13 where refitted in this manner following the outline by McCluskey $et$ $al$.\cite{McCluskey_2020} The evidence terms derived from these fits are shown Figure \ref{PNR14} (c) and Table \ref{table3}. Figure \ref{PNR14} (a) also shows the normalized $\chi^{2}$ values as a function of number of parameters and model number. It is clear that Model 8 and 9 are very close, together with 8 being slightly better than 9. There were several bi-modal correlations in the fit in particular regarding the graphene thickness and the graphene and top Ni interface roughnesses. Some of the cleanest examples of which are shown in Figure \ref{PNR15}. This shows how crucial the inclusion of the secondary information is in determining the best model as it effectively allows the selection of the right node. The result from the nested sampler, following that the biggest value represents the greatest evidence and is the most probable, i.e. model 9.


In summary, the decision on the best PNR model was taken by considering the information obtained from the complementary techniques (SEM, Raman spectroscopy and XMCD). Therefore, model 9 was used to estimate the magnetic moment induced in graphene in the rotated-domain graphene/Ni sample. The same model was then used to fit the PNR data of the other systems; epitaxial graphene/Ni and graphene/Ni$_{9}$Mo$_{1}$ samples.

\bibliography{library.bib}